\documentclass[12pt]{iopart}

\usepackage{iopams}
\usepackage{graphics}
\usepackage{graphicx}
\usepackage{epsfig}
\usepackage{rotating}
\usepackage[latin1]{inputenc}
\usepackage{hyperref}

\begin{document}

\title[Traffic jams in two-lane transport]{Traffic jams induced by rare switching events in two-lane transport}

\author{Tobias Reichenbach, Erwin Frey, and Thomas Franosch}

\address{Arnold Sommerfeld Center for Theoretical Physics
(ASC) and
  Center for NanoScience (CeNS), Department of Physics,
  Ludwig-Maximilians-Universit\"at M\"unchen, Theresienstrasse 37,
  D-80333 M\"unchen, Germany}
\ead{tobias.reichenbach@physik.lmu.de}
\begin{abstract}
We investigate a model for driven exclusion processes where
internal states are assigned to the particles. The latter account
for diverse situations, ranging from spin states in spintronics to
parallel lanes in intracellular or vehicular traffic. Introducing
a coupling between the internal states by allowing particles to
switch from one to another induces an intriguing polarization
phenomenon. In a mesoscopic scaling, a rich stationary regime for
the density profiles is discovered, with localized domain walls in
the density profile of one of the internal states being feasible.
We derive the shape of the density profiles as well as resulting
phase diagrams analytically by a mean-field approximation and a
continuum limit. Continuous as well as discontinuous lines of
phase transition emerge, their intersections induce multicritical
behavior.
\end{abstract}

\maketitle

\section{Introduction}

Non-equilibrium critical phenomena arise in a broad variety of
systems, including  non-equilibrium growth models~\cite{barabasi},
percolation-like processes~\cite{deutscher}, kinetic Ising
models~\cite{droz-1989-39}, diffusion limited chemical
reactions~\cite{mattis-1998-70}, and driven diffusive
systems~\cite{SchmittmannZia}. The latter provide models for
transport processes ranging from biological systems, like the
motion of ribosomes along a \emph{m}-RNA
chain~\cite{macdonald-1968-6} or processive motors walking along
cytoskeletal filaments~\cite{Howard,hirokawa-1998-279}, to
vehicular traffic~\cite{helbing-2001-73,chowdhury-2000-329}. In
this work, we focus on the steady-state properties of such
one-dimensional transport models, for which the Totally Asymmetric
Simple Exclusion Process (TASEP) has emerged as a paradigm (for
reviews  see e.g.~\cite{Derrida,Mukamel,Schutz}). There, particles
move unidirectionally from  left to  right on a one-dimensional
lattice, interacting through on-site exclusion. The entrance/exit
rates at the open left/right boundary control the system's
behavior; tuning them, one encounters different non-equilibrium
phases for the particle densities~\cite{krug-1991-76}.

Intense theoretical research has been devoted to the
classification of such non-equilibrium phenomena. For example,
within the context of reaction-diffusion systems, there is strong
evidence that phase transitions from an active to an absorbing
state can be characterized in terms of only a few universality
classes, the most important being the one of directed percolation
(DP)~\cite{odor-2004-76}. To search for novel critical behavior,
fruitful results have been obtained by coupling two
reaction-diffusion systems~\cite{tauber-1998-80,noh-2005-94}, each
undergoing the active to absorbing phase transition. Due to the
coupling, the system exhibits a multicritical point with unusual
critical behavior.

We want to stress that already in equilibrium physics seminal
insights have been gained by  coupling identical systems. For
instance, spin-ladders incorporate several Heisenberg spin chains
\cite{dagotto-1995-271}. There, quantum effects lead to a
sensitive  dependence on the chain number: for even ones a finite
energy gap between the ground state and the lowest excitation
emerges whereas gapless excitations dominate the low-temperature
behavior if the number of spin chains is odd.

In this work, we generalize the Totally Asymmetric Exclusion
Process (TASEP) in a way that particles  possess two internal
states; we have recently published a short account of this work in Ref.~\cite{reichenbach-2006-97}. Allowing particles to occasionally
switch from one internal
state to the other induces a coupling between the latter; indeed,
the model may alternatively be regarded as two coupled TASEPs.
When independent, each of them separately undergoes
boundary-induced phase transitions~\cite{krug-1991-76}. The
coupling is expected to induce novel phenomena, which are the
subject of the present work.

 \begin{figure}[htbp]
\begin{center}
\includegraphics[scale=1]{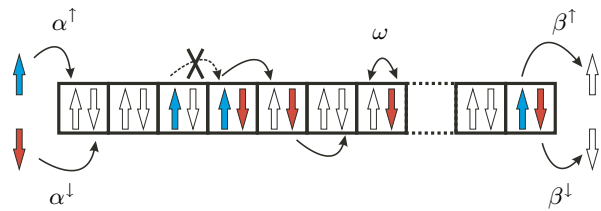}
\caption{(Color online) Illustration of an exclusion model with
two internal states, adopting the language of spin transport.
Particles in states $\uparrow$ ($\downarrow$) enter with rates
$\alpha^{\uparrow}$ ($\alpha^\downarrow$), move unidirectionally
to the right within the lattice, may flip at rate $\omega$ and
leave the system at rates $\beta^{\uparrow}$ ($\beta^\downarrow$),
always respecting Pauli's exclusion principle.
\label{cartoon_spin}}
\end{center}
\end{figure}

Exclusion is introduced by allowing multiple occupancy of lattice
sites only if particles are in different internal states. Viewing
the latter as spin-$1/2$ states, i.e spin-up~($\uparrow$) and
spin-down~($\downarrow$), this  directly translates into Pauli's
exclusion principle; see Fig.~\ref{cartoon_spin}.  Indeed, the
exclusion process presented in this work may serve as a model for
semiclassical transport in mesoscopic quantum systems
\cite{zutic-2004-76}, like hopping transport in chains of quantum
dots in the presence of an applied field~\cite{hahn-1998-73}. Our
model incorporates the quantum nature of the particles through
Pauli's exclusion principle, though phase coherence is ignored. A
surprising analogy to a simple spintronics scheme, the Datta-Das
spin field-effect transistor~\cite{zutic-2004-76}, holds. There,
electrons move unidirectionally through a ferromagnetic metal or a
semiconductor. The polarization of the electrons is controllable
by a source for spin injection, a drain for spin extraction as
well as a gate in the form of  a tunable magnetic field that
controls the strength of spin precession. In our model, this is
mimicked by considering the spin-flip rate as a control parameter.

The model is potentially relevant within biological contexts, as
well. In intracellular traffic~\cite{Howard,Hinsch}, 
molecular motors walking on parallel filaments may  detach from
one lane and attach on another, resulting in an effective
switching between the lanes. In our model, identifying the two
internal states with different lanes, one recovers a transport
model on two lanes with simple site exclusion. In the same way,
the system presented in this work serves as a highly simplified cartoon model  of   
multi-lane highway traffic taking lane switching into
account~\cite{helbing-2001-73, chowdhury-2000-329}.

\begin{figure}
\begin{center}
\includegraphics[scale=1]{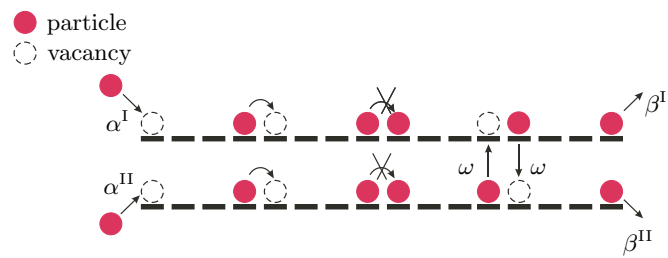}
\caption{Illustration of the two-lane interpretation. We label the
upper lane as lane $\textnormal{I}$ and the lower one as lane
$\textnormal{II}$. They possess individual entering rates,
$\alpha^\textnormal{\scriptsize I}$ resp. $\alpha^\textnormal{\scriptsize II}$ as well as exiting
rates, $\beta^\textnormal{\scriptsize I}$ resp.
$\beta^\textnormal{\scriptsize II}$.
\label{cartoon_two_lane}}
\end{center}
\end{figure}

Significant insight into  multi-lane traffic has
been achieved (see Ref.~\cite{helbing-2001-73, chowdhury-2000-329} and references therein).
In particular, novel phases have been discovered in the case of
indirect coupling, i.e. the velocity of the particles depends on
the configuration on the neighboring lane~\cite{popkov-2001-64,
popkov-2003-112,popkov-2004-37}. Recently, models have been
presented that allow particles to switch between  lanes, and the
transport properties have in part been rationalized in terms of an
effective single lane
TASEP~\cite{pronina-2004-37,mitsudo-2005-38,pronina-2006-372}. There, the case of strong coupling has been investigated: the timescale of lane switching events is the same as of forward hopping. In
our model, we explicitly want to  ensure a competition between the
boundary processes and the switching between the internal states.
We therefore employ a mesoscopic scaling, i.e. we consider the
case where the switching events are rare as compared to forward hopping. This is the  situation  encountered in intracellular traffic \cite{Howard} where motors nearly exclusively remain on one lane and  switch only very rarely. 
In the context of spin transport, it corresponds to the case where forward hopping occurs much faster than spin precession (weak external magnetic field).

The outline of the present paper is the following. In
Sec.~\ref{model} we introduce the model in the context of spin
transport as well as two-lane traffic. Symmetries and currents are
discussed, which play a key role in the following analysis.
Section~\ref{mean-cont} describes in detail the mean-field
approximation and the differential equations for the densities
obtained therefrom through a continuum limit. The mesoscopic
scaling is motivated and introduced, the details of the analytic
solution for the spatial density profiles being condensed in
\ref{app_first_order}.  We obtain the generic form of the
density profiles in Sec.~\ref{gen_dens}, and compare our analytic
results  to stochastic simulations. We find that they agree excellently, suggesting
the exactness of our analytic approach in the limit of large systems.
As our main result, we encounter the polarization phenomenon, where the density profiles in the stationary non-equilibrium state exhibit localized `shocks'. Namely, the density of one spin state changes abruptly from  low to high  density. The origin of this phenomenon is rationalized in terms of
singularities in coupled differential equations. We partition the
full parameter space into three distinct regions, and observe a
delocalization transition. The methods to calculate the phase
boundaries analytically are developed simultaneously.
Section~\ref{stoch_sim} presents details on the stochastic
simulations which we have carried out to corroborate our analytic
approach. The central result of this work is then addressed in
Sec.~\ref{phase_diag}, where two-dimensional analytic phase
diagrams are investigated. Our analytic approach identifies the phases where the polarization phenomenon occurs, as well as the continuous and discontinuous transitions that separate the  phases. The nature of the transitions is explained by the injection/extraction limited current which 
is conserved  along the track. As a second remarkable feature of the model, we uncover multi-critical points, i.e. points where two lines of phase boundaries intersect or the nature of a  phase transition changes from discontinuous to a continuous one. Although multi-critical point are well-known in equilibrium statistical mechanics, a fundamental description for such a behavior 
for  systems  driven far from equilibrium still constitutes a major challenge.  
 A brief summary and outlook
concludes this work.

\section{The model}
   \label{model}

In this section, we describe our model in terms of spin transport
as well as two-lane traffic. Though we will preferentially use the
language of spins in the subsequent sections, the two-lane
interpretation is of no lesser interest, and straightforwardly
obtained. Furthermore, we introduce two symmetries which are
manifest on the level of the dynamical rules.

\subsection{Dynamical rules}
\label{dynamics}

We consider hopping transport on a one-dimensional lattice,
composed of $L$ sites, with open boundaries, see
Fig.~\ref{cartoon_spin}. Particles possess internal states, which
we restrict to two different kinds; adopting a spin notation, they
are referred to as spin-up ($\uparrow$) and spin-down
($\downarrow$). They enter at the left boundary at rates
$\alpha^\uparrow$ resp. $\alpha^\downarrow$, and move
unidirectionally from left to the right through the lattice. The
timescale is fixed by putting the rate for these hopping events to
unity. Within the bulk, particles may also flip their spin state, from
spin-up to spin-down and back, at  rate $\omega$. Finally, having
reached the right boundary, particles may exit the system at rates
$\beta^\uparrow$ resp. $\beta^\downarrow$, depending on their spin
state. We allow all of these processes only under the constraint
of Pauli's exclusion principle, meaning that every lattice site
may at most be occupied by one particle of a given state. Spin-up
and spin-down thus may simultaneously occupy the same site,
however two particles with identical spin polarization cannot
share a lattice site. In summary, our dynamical rules are the
following:
\begin{itemize}
\item[(i)] at site $i=1$ (left boundary), particles with
spin-up (spin-down) may enter at rate $\alpha^\uparrow$
($\alpha^\downarrow$)
\item[(ii)] at site $i=L$ (right boundary), particles
with spin-up (spin-down) leave the lattice at rate
$\beta^\uparrow$ ($\beta^\downarrow$)
\item[(iii)] particles may hop at unit
rate from site $i-1$ to the neighboring site $i$ for $i \in
\{2,\ldots,L\}$, i.e. within bulk
\item[(iv)] within bulk, particles can flip their spin
state with rate $\omega$, i.e. spin-up turns into spin-down and
vice versa
\end{itemize}
always respecting Pauli's exclusion principle. Processes (i) to
(iii) constitute the Totally Asymmetric Simple Exclusion Processes
(TASEP) for the two different states separately, while rule (iv)
induces a coupling between them. Indeed, when the spin-flip rate
$\omega$ vanishes, we recover the trivial situation of two
independent TASEPs, while we will show that a proper treatment of
$\omega$ through a mesoscopic scaling induces nontrivial effects.

\subsection{Two-lane interpretation}

Having introduced our model in the language of semi-classical spin
transport, where Pauli's exclusion principle is respected while
phase coherence completely ignored, we now want to show that it
also describes transport with site exclusion on two parallel
lanes. As schematically drawn in Fig.~\ref{cartoon_two_lane}, we
consider two parallel lanes, each consisting of $L$ sites, labeled
as upper lane (I) and lower lane (II). They are identified with
the internal states of the particles considered before: a particle
with spin-up (spin-down) now corresponds to a particle on lane I
(lane II). The processes (i) and (ii) describe entering of
particles at lane I (II) at rate
$\alpha^\textnormal{\scriptsize I}\equiv\alpha^\uparrow$
($\alpha^\textnormal{\scriptsize II}\equiv\alpha^\downarrow$) and exiting of lane I
(II) at  rate $\beta^\textnormal{\scriptsize I}\equiv\beta^\uparrow$
($\beta^\textnormal{\scriptsize II}\equiv\beta^\downarrow$). Due to (iii), particles
hop unidirectionally to the right on each individual lane; at rate
$\omega$, they may switch from lane I to II and back. Pauli's
exclusion principle translates into simple site exclusion: all the
above processes are allowed under the constraint of admitting at
most one particle per site. Again, we clearly observe that it is
process (iv) that couples two TASEPs, namely the ones on each
individual lane, to each other.

\subsection{Symmetries}  

Already on the level of the dynamical rules (i)-(iv) presented
above, two symmetries are manifest that will prove helpful in the
analysis of the system's behavior. We refer to  the absence of
particles with certain state as holes with the opposite respective
state~\footnote{The convention to flip the spin simultaneously is
natural in the language of solid-state physics. In the context of
two-lane traffic, it appears more natural to consider vacancies
moving on the \emph{same} lane in the reverse direction.}.
Considering their motion, we observe that the dynamics of the
holes is governed by the identical rules (i) to (iv), with
``left'' and ``right'' interchanged, i.e. with a discrete
transformation of sites $i\leftrightarrow L-i$ as well as rates
$\alpha^{\uparrow, \downarrow}\leftrightarrow \beta^{\downarrow,
\uparrow}$. The system thus exhibits a \emph{particle-hole
symmetry}. Even more intuitively, the two states behave
qualitatively identical. Indeed, the system remains invariant upon
changing spin-up to spin-down states and vice versa with a
simultaneous interchange of $\alpha^\uparrow\leftrightarrow
\alpha^\downarrow$ and $\beta^\uparrow\leftrightarrow
\beta^\downarrow$, constituting a \emph{spin symmetry} (in terms
of the two-lane interpretation, it translates into a \emph{lane
symmetry}).\\ When analyzing the system's behavior in the
five-dimensional phase space, constituted of the entrance and exit
rates $\alpha^{\uparrow, \downarrow},~\beta^{\uparrow,
\downarrow}$ and $\omega$, these symmetries  allow to connect
different regions in phase space, and along the way to simplify
the discussion.

\section{Mean-field equations, currents, and the continuum limit}

\label{mean-cont}

In this section, we shall make use of the  dynamical rules introduced
above to set up a quantitative description for the densities and
currents in the system. Within  a mean-field approximation, their
time evolution is expressed through one-point functions only,
namely the average occupations of a lattice site. Such mean-field
approximations have been successfully applied to a variety of driven diffusive systems, see e.g. Ref.~\cite{Mukamel}. We focus on the properties of the non-equilibrium steady state, which results from
boundary processes (entering and exiting events) as well as bulk
ones (hopping and spin-flip events). Both types of processes
compete if their time-scales are comparable; we ensure this
condition by introducing a \emph{mesoscopic scaling} for the spin
flip rate~$\omega$. Our focus is on the limit of large system
sizes $L$, which is expected to single out distinct phases. To
solve the resulting equations for the densities and currents,  a
continuum limit is then justified, and it suffices to consider the
leading order in the small parameter, viz. the ratio of the
lattice constant to system size. Such a mesoscopic scaling has  been already
successfully used in~\cite{parmeggiani-2003-90,parmeggiani-2004-70} in the
context of TASEP coupled to Langmuir dynamics.\\
 
\subsection{Mean field approximation and currents}

Let $n_i^{\uparrow}(t)$ resp. $n_i^{\downarrow}(t)$ be the
fluctuating occupation number of site~$i$ for spin-up resp.
spin-down state, i.e. $n_i^{\uparrow,\downarrow}(t)=1$ if
 this site is occupied at time $t$ by a particle with
the specified spin state and $n_i^{\uparrow,\downarrow}(t)=0$
otherwise. Performing ensemble averages, the expected occupation,
denoted by $\rho_i^\uparrow(t)$ and $\rho_i^\downarrow(t)$, is
obtained.
 Within a mean-field
approximation,  higher order
 correlations
between the occupation numbers are neglected, i.e. we impose the factorization approximation
\begin{equation} \label{decoupling}
\langle n_i^r(t) n_j^s(t)\rangle=\rho_i^r(t)\rho_j^s(t) \,; \quad
r,s\in\{\uparrow,\downarrow\}\, .
\end{equation}

Equations of motion for the densities can by  obtained via
\emph{balance equations:} The time-change of the density at a
certain site is related to appropriate currents. The spatially
varying \emph{spin current} $j_i^{\uparrow}(t)$  quantifies  the
rate at which particles of spin state $\uparrow$ at site $i-1$ hop
to the neighboring site $i$. Within the  mean-field approximation,
Eq. (\ref{decoupling}), the current is expressed in terms of
densities as
\begin{equation}
j_i^{\uparrow}(t)=\rho_{i-1}^\uparrow(t)[1-\rho_i^\downarrow(t)]
\, , \qquad i \in \{2,\ldots,L\} \, ,\label{curr_a}
\end{equation}
and similarly for the current $j_i^{\downarrow}(t)$. The sum
yields the total \emph{particle current} $J_i(t)\equiv
j_i^\uparrow(t)+j_i^\downarrow(t)$.
 Due to the spin-flip process
(iv), there also exists a \emph{leakage current}
$j_i^{\uparrow\downarrow}(t)$ from spin-up state to spin-down
state. Within mean-field
\begin{equation}\label{curr_updown}
j_i^{\uparrow\downarrow}(t)=\omega\rho_i^\uparrow(t)[1-\rho_i^\downarrow(t)]\,
,
\end{equation}
and similarly for the leakage current $j_i^{\downarrow\uparrow}(t)$ from
spin-down to spin-up state. Now, for $i \in \{2,\ldots, L-1\}$ we
can use balance equations to obtain the time evolution of the
densities,
\begin{eqnarray}
\frac{\textnormal{d}}{\textnormal{d} t} \rho_i^\uparrow(t)=&j_i^{\uparrow}(t)
-j_{i+1}^{\uparrow}(t)+j_i^{\downarrow\uparrow}(t)-j_i^{\uparrow\downarrow}(t)
\, . \label{occ_curr}
\end{eqnarray}
This constitutes an exact relation. Together with the mean field
approximation for the currents, Eqs. (\ref{curr_a},
\ref{curr_updown}), one obtains a set of closed equations for the local densities
\begin{eqnarray}
\frac{\textnormal{d}}{\textnormal{d} t}
\rho_i^\uparrow(t)=~&\rho_{i-1}^\uparrow(t)[1-\rho_i^\uparrow(t)]-\rho_{i}^\uparrow(t)[1-\rho_{i+1}^\uparrow(t)]+\omega\rho_i^\downarrow(t) -\omega\rho_i^\uparrow(t) \, .
\label{occ_num}
\end{eqnarray}
At the boundaries of the track, the corresponding expressions involve also the
entrance and exit events, which are again treated  in the spirit of
a  mean-field approach
\begin{eqnarray}
\frac{\textnormal{d}}{\textnormal{d} t}
\rho_1^\uparrow(t)=~&\alpha^\uparrow[1-\rho_1^\uparrow(t)]
-\rho_{1}^\uparrow(t)[1-\rho_{2}^\uparrow(t)] +\omega\rho_1^\downarrow(t)-\omega\rho_1^\uparrow(t)\, ,
\label{bound_1}
\\
\frac{\textnormal{d}}{\textnormal{d} t}
\rho_L^\uparrow(t)=~&\rho_{L-1}^\uparrow(t)[1-\rho_L^\uparrow(t)]
-\beta^\uparrow\rho_{L}^\uparrow(t)+\omega\rho_L^\downarrow(t)-\omega\rho_L^\uparrow(t)
\label{bound_N}\, .
\end{eqnarray}
Due to the spin symmetry, i.e. interchanging $\uparrow$ and
$\downarrow$, an analogous set of equations hods for the time evolution
of the density of particles with spin-down state. \\

In the stationary state, the densities
$\rho^{\uparrow(\downarrow)}_i(t)$ do not depend on time $t$, such
that  the time derivatives  in
Eqs.~(\ref{occ_num})-(\ref{bound_N}) vanish. Therefrom, we
immediately derive the spatial conservation of the particle
current: Indeed, summing Eq.~(\ref{occ_curr}) with the
corresponding equation for the density of spin-down states yields
\begin{equation}
J_i=J_{i+1}\, , \qquad i \in \{2,\ldots,L-1\}\, ,
\end{equation}
such that the particle current does not depend on the spatial
position $i$. Note that this does not apply to the individual spin
currents,  they \emph{do} have a spatial dependence arising from
the leakage currents.\\

In a qualitative discussion, let us  now anticipate the effects
that arise from the non-conserved individual spin currents  as
well as from the conserved particle current. The latter has its
analogy in TASEP,  where the particle current is spatially
conserved as well. It leads to two distinct regions in the
parameter space: one where the current is determined by the left
boundary, and the other where it is controlled by the right one.
Both regions are connected by the discrete particle-hole symmetry.
Thus, in general, discontinuous phase transitions arise when crossing the
border from one region to the other. In our model, we will find
similar behavior: the particle current is either determined by the
left or by the right boundary. Again, both regions are connected
by the discrete particle-hole symmetry, such that we expect
discontinuous phase transitions at the border between both. Except
for a small, particular region in the parameter space, this
behavior is captured quantitatively by the mean-field approach and
the subsequent analysis, which is further corroborated by stochastic
simulations. The phenomena linked to the particular region  will
be presented elsewhere~\cite{reichenbach-deloc}.\\ On the other
hand, the non-conserved spin currents  may be compared to the
current in TASEP coupled to Langmuir kinetics; see Refs.~\cite{parmeggiani-2003-90,parmeggiani-2004-70}. Due to attachment
and detachment processes, the in-lane current is only weakly
conserved, allowing for a novel phenomena, namely phase separation
into a low-density and a high-density region separated by a
localized domain wall. The transitions to this phase are
continuous considering the domain wall position $x_w$ as the order
parameter. In our model, an analogous but even more intriguing phase will appear as well, with
continuous transitions being possible.

\subsection{Mesoscopic scaling and the  continuum limit}

\subsubsection{Mesoscopic scaling.}

Phases and corresponding phase transitions are expected to emerge
in the limit of large system size, $L\rightarrow\infty$, which
therefore  constitutes the focus of this work. We expect
interesting phase behavior  to arise from the coupling of spin-up
and spin-down states via spin-flip events, in addition to the
entrance and exit processes.   Clearly, if spin-flips occur on a
fast time-scale, comparable to the hopping events, the spin degree
of freedom is relaxed, such that the system's behavior is
effectively the one of a TASEP. Previous work on related two-lane
models~\cite{mitsudo-2005-38, pronina-2004-37} focused on the
physics in that situation. In this work, we want to highlight the
dynamical regime where coupling through spin-flips is present,
however not sufficiently strong to relax the system's internal
degree of freedom. In other words, we consider physical situations
where  spin-flips occur on the same
time-scale as the entrance/exit processes. Defining the
\emph{gross} spin-flip rate $\Omega=\omega L$ yields a measure of
how often a particle flips its spin state while traversing the
system. To ensure competition between spin-flips with boundary
processes, a \emph{mesoscopic scaling} of the rate $\omega$ is
employed by keeping $\Omega$ fixed, of the same order as the
entrance/exit rates, when the number of lattice sites becomes
large~$L\rightarrow\infty$.

\subsubsection{Continuum limit and first order approximation.}

The total length of the lattice will be fixed to unity and one may
define consistently the lattice constant $\epsilon=1/L$. In the
limit of large systems $\epsilon\rightarrow 0$,  a \emph{continuum
limit} is anticipated. We introduce continuous functions
$\rho^\uparrow(x)$ resp. $\rho^\downarrow(x)$ through
$\rho^\uparrow(x_i)=\rho^\uparrow_i$ resp.
$\rho^\downarrow(x_i)=\rho^\downarrow_i$ at the discrete points
$x_i=i\epsilon$. Expanding these to first order in the lattice
constant,
 \begin{equation}
\rho^{\uparrow(\downarrow)}(x_{i\pm1})=\rho^{\uparrow(\downarrow)}(x_i\pm\epsilon)
=\rho^{\uparrow(\downarrow)}(x_i)\pm\epsilon\partial_x\rho^{\uparrow(\downarrow)}(x_i)
\, ,
\end{equation}
the difference equations (\ref{occ_num})-(\ref{bound_N}) turn into
differential equations. Observing  that $\omega=\epsilon\Omega$ is
already of order $\epsilon$, we find that the zeroth order of
Eq.~(\ref{occ_num}) vanishes, and the first order in $\epsilon$
yields
\begin{equation}
[2\rho^\uparrow(x)-1]\partial_x\rho^\uparrow(x)
+\Omega\rho^\downarrow(x)-\Omega\rho^\uparrow(x)=0  \label{cont_a}
\, .
\end{equation}
Similarly, the same manipulations  for $\rho^\downarrow$ yield
\begin{equation}
[2\rho^\downarrow(x)-1]\partial_x\rho^\downarrow(x)
+\Omega\rho^\uparrow(x)-\Omega\rho^\downarrow(x)=0 \, .
\label{cont_b}
\end{equation}
The expansion of  Eqs. (\ref{bound_1}) and (\ref{bound_N}) in
powers of $\epsilon$, yields in  zeroth order
\begin{eqnarray}
\rho^\uparrow(0) & =&~\alpha^\uparrow\, , \qquad
 \rho^\uparrow(1)=
1-\beta^\uparrow\, , \nonumber\\ \rho^\downarrow(0)&=&
~\alpha^\downarrow\, , \qquad
\rho^\downarrow(1)=~1-\beta^\downarrow \, , \label{bound_cond}
\end{eqnarray}
which impose \emph{boundary conditions}. Since two boundary
conditions are enough to specify a solution of the coupled first
order differential equations, the system is apparently
over-determined.
  Of course, the full analytic solution,
i.e. where all orders in $\epsilon$ are incorporated, will  be only
\emph{piecewise} given by the first-order approximation, Eqs.~(\ref{cont_a})-(\ref{bound_cond}). Between these branches, the
solution will depend on higher orders of $\epsilon$, therefore,
these intermediate regions  scale with order $\epsilon$ and
higher. They vanish in the limit of large systems,
$\epsilon\rightarrow 0$, yielding  \emph{domain walls} or
\emph{boundary layers}.\\ Let us explain the latter terms. At the
position of a domain wall, situated in bulk, the density changes
its value discontinuously,  from one of a low-density region to
one of a high-density.  Boundary layers are pinned to the
boundaries of the system. There as well, the density changes
discontinuously: from a value that is given by the corresponding boundary
condition to that of a low- or high-density region which is imposed by the opposite boundary .

\subsubsection{Symmetries and currents revisited.}

In the following, we reflect important properties of the system,
symmetries and currents, on the level of the first-order
approximation,  Eqs.~(\ref{cont_a})-(\ref{bound_cond}).  The explicit
solution of the latter can be found in \ref{app_first_order}.

The particle-hole symmetry, already inferred   from the dynamical
rules, now takes the form
\begin{eqnarray}
\rho^{\uparrow(\downarrow)}(x)\leftrightarrow&~
1-\rho^{\uparrow(\downarrow)}(1-x)\, , \cr
\alpha^{\uparrow(\downarrow)}\leftrightarrow&~\beta^{\uparrow(\downarrow)}
\, .
\end{eqnarray}
Interchanging~$\uparrow$ and~$\downarrow$ in the  densities as
well as the in- and outgoing rates yields the spin symmetry,
\begin{eqnarray}
\rho^\uparrow(x)&\leftrightarrow \rho^\downarrow(x) \, ,\cr
\alpha^\uparrow&\leftrightarrow \alpha^\downarrow \, ,\cr
\beta^\uparrow&\leftrightarrow \beta^\downarrow  \, .
\end{eqnarray}

The individual spin currents as well as the particle current have
been anticipated to provide further understanding of the system's
behavior. In the continuum limit the zeroth order of the spin
currents  is found to be
$j^{\uparrow(\downarrow)}(x)=\rho^{\uparrow(\downarrow)}(x)\big[1-\rho^{\uparrow(\downarrow)}(x)\big]$,
such that Eqs.~(\ref{cont_a}), (\ref{cont_b}) may be written in
the form
\begin{equation}
\partial_x j^\uparrow=\Omega[\rho^\downarrow-\rho^\uparrow]~
,\quad \partial_x
j^\downarrow=\Omega[\rho^\uparrow-\rho^\downarrow]\, .
\label{cont_curr}
\end{equation}
The terms on the right-hand side, arising from the spin-flip
process~(iv), are seen to violate the spatial conservation of the
spin currents. However, due to the mesoscopic scaling of the spin
flip rate~$\omega$, the leakage currents between the spin states
are only weak, see Eq. (\ref{curr_updown}), locally tending to zero when $\epsilon\rightarrow
0$, such that the spin currents vary \emph{continuously} in space.
This finding imposes a condition for the transition from one
branch of first-order solution to another, as described above:
such a transition is only allowed when the corresponding spin
currents are continuous at the transition point, thus singling out
distinct positions for a possible transition.\\ Finally, summing
the two equations in Eq.~(\ref{cont_curr}) yields the spatial
conservation of the particle current: $\partial_x J=0$.

\section{Partition of the parameter space and the generic density behavior}

\label{gen_dens}

The parameter space of our model, spanned by the five rates $\alpha^{\uparrow, \downarrow},~\beta^{\uparrow,
\downarrow}$, and $\Omega$, is of high dimensionality. However, in this section, we show that it can be  decomposed into only   three basic distinct regions: the
maximal-current region (MC) as well as the injection-limited (IN) and the
extraction-limited one (EX). While trivial phase behavior occurs
in the MC region, our focus is on the IN  and EX region
(connected by particle-hole symmetry), where a striking
polarization phenomenon occurs. The generic phase behavior in
these regions is derived, exhibiting this effect.

\subsection{Effective rates}

\begin{figure}
\begin{center}
\includegraphics[scale=1]{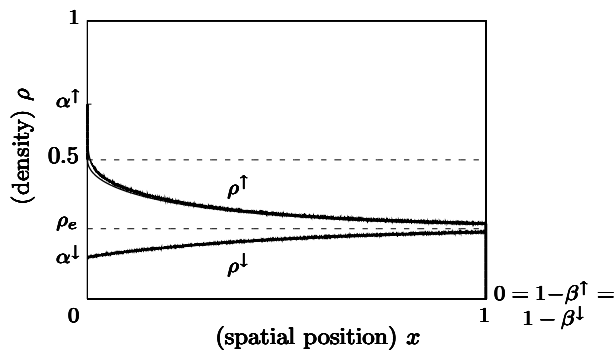}
\caption{Illustration of the effective rates. The right boundary
is ``open'', such that only  the capacity of the bulk and the
entrance rates limit the spin currents. The injection rate
$\alpha^\uparrow>\frac{1}{2}$ effectively acts as $\frac{1}{2}$.
The analytic predictions correspond to the solid lines, the
results from stochastic simulations for $L=10000$ are indicated by
the wiggly line. With increasing spatial position,  the densities
approach a common value $\rho_e$. The parameters used are
$\alpha^\uparrow=0.7,~\alpha^\downarrow=0.15,~\Omega=0.5$.\label{eff_rates}}
\end{center}
\end{figure}
The entrance and exit rates as well as the carrying capacity of
the bulk impose restrictions on the particle current. For example, the capacity
of the bulk limits the individual spin currents $j^{\uparrow(\downarrow)}$ to a maximal values
of $1/4$. The latter  occurs  at a density of $1/2$, as seen
from the previous result
$j^{\uparrow(\downarrow)}=\rho^{\uparrow(\downarrow)}[1-\rho^{\uparrow(\downarrow)})]$. To
illustrate  the influence of the injection and extraction rates,
we first consider an ``open''  right boundary  i.e. $\beta^\uparrow=\beta^\downarrow=1$. Particles then
leave the system unhindered, such that only the entrance rates may
limit the particle current. Provided one of these rates, say
$\alpha^\uparrow$, exceeds the value $1/2$, the current of the
corresponding state ($\uparrow$) is limited by the capacity of the
bulk to a value of $1/4$ in the vicinity of the left boundary.  A boundary layer thus forms in the density
profile of spin-up state at the left boundary, connecting the
value of the injection rate $\alpha^\uparrow$ to the value $1/2$. Up to this
boundary layer, the density profile $\rho^\uparrow(x)$ is
identical to the one where $\alpha^\uparrow$ takes a value of
$1/2$, c.f. Fig.~\ref{eff_rates}. Similar reasoning holds for the
extraction rates $\beta^{\uparrow(\downarrow)}$. They as well
behave effectively as $1/2$ when exceeding this value.  To treat
these findings properly, we introduce the \emph{effective rates}
\numparts
\begin{eqnarray} 
\alpha_{\textnormal{\scriptsize eff}}^{\uparrow(\downarrow)}&=&\textnormal{ min}\Big[\alpha^{\uparrow(\downarrow)}, \frac{1}{2}\Big]\,
,\\
\beta_{\textnormal{\scriptsize eff}}^{\uparrow(\downarrow)}&=&\textnormal{ min}\Big[\beta^{\uparrow(\downarrow)},\frac{1}{2}\Big]
\, . 
 \end{eqnarray}
 \endnumparts
The system's bulk  behavior will only depend on them, and, in
particular, remain unaffected when a  rate is varied at values
exceeding $1/2$.

\subsection{Injection-limited, extraction-limited, and maximal current region}

Equipped with these results, in the case of an ``open'' right
boundary, the spin currents in the vicinity of the left boundary
are given by
$j^\uparrow=\alpha_\textnormal{\scriptsize eff}^\uparrow(1-\alpha_\textnormal{\scriptsize eff}^\uparrow)$
resp.
$j^\downarrow=\alpha_\textnormal{\scriptsize eff}^\downarrow(1-\alpha_\textnormal{\scriptsize eff}^\downarrow)$,
resulting in a particle current $J_\textnormal{\scriptsize IN}$ imposed by the
injection rates:
$J_\textnormal{\scriptsize IN}=\alpha_\textnormal{\scriptsize eff}^\uparrow(1-\alpha_\textnormal{\scriptsize eff}^\uparrow)
+\alpha_\textnormal{\scriptsize eff}^\downarrow(1-\alpha_\textnormal{\scriptsize eff}^\downarrow)$.
The analogous relations, with the injection and extraction rates
interchanged, hold for the case of an ``open'' left boundary,
$\alpha^\uparrow=\alpha^\downarrow=1$. The particle current is
then controlled by the right boundary:
$J_\textnormal{\scriptsize EX}=\beta_\textnormal{\scriptsize eff}^\uparrow(1-\beta_\textnormal{\scriptsize eff}^\uparrow)
+\beta_\textnormal{\scriptsize eff}^\downarrow(1-\beta_\textnormal{\scriptsize eff}^\downarrow)$. In
general, depending on which imposes the stronger restriction,
either the left or the right boundary limits the particle current:
$J\leq\textnormal{ min}(J_\textnormal{\scriptsize IN},J_\textnormal{\scriptsize EX})$. Indeed,
$J=\textnormal{ min}(J_\textnormal{\scriptsize IN},J_\textnormal{\scriptsize EX})$ holds except for an
anomalous situation, where the current is lower than this value \footnote{This situation arises in a certain neighborhood of the multicritical points $\mathcal{B}$, discussed in Sec.~\ref{phase_diag}}. Depending on which of both cases applies, two
complementary regions in phase space are distinguished:
$J_\textnormal{\scriptsize IN}<J_\textnormal{\scriptsize EX}$ is termed \emph{injection-limited
region} (IN), while $J_\textnormal{\scriptsize IN}>J_\textnormal{\scriptsize EX}$ defines the
\emph{extraction-limited region} (EX). Since they are connected by
 discrete particle-hole symmetry, we expect \emph{discontinuous
phase transitions} across the border between both, to be referred
as \emph{IN-EX boundary}. 

Right at the IN-EX boundary, the system exhibits coexistence of low- and high-density phases, separated by domain walls. Interestingly, this phase coexistence emerges on \emph{both} lanes (states), which may be seen as follows. Recall that a domain wall concatenates a region of low and another of high density. However, while the densities exhibit a discontinuity, the spin currents must be continuous. In other words, the spin currents, and therefore the particle currents, imposed by the left and right boundary must match each other. This yields the condition $J_\textnormal{\scriptsize IN}=J_\textnormal{\scriptsize EX}$, which is nothing but the relation describing the IN-EX boundary. Actually, what we have shown with this argument is that domain walls on both lanes (states) are \emph{at most} feasible at the IN-EX boundary. However, it turns out that there, they do indeed form, and are delocalized. We refer to our forthcoming publication~\cite{reichenbach-deloc} for a detailed discussion of this phenomenon. Away from the IN-EX boundary, it follows that at most on one lane (state) a domain wall may appear.

When both entrance rates $\alpha^\uparrow,~\alpha^\downarrow$ as
well as both exit rates $\beta^\uparrow,~\beta^\downarrow$ exceed
the value  $1/2$, the particle current is limited by neither
boundary, but only through the carrying capacity of the bulk,
restricting it to twice the maximal value $1/4$ of the individual
spin currents: $J=1/2$. The latter situation therefore constitutes
the \emph{maximal current region}~(MC).

\subsection{The generic state of the densities}

As we have seen in the previous section, particularly simple density 
profiles emerge in the MC region. There, up to boundary layers, the density 
profiles remain constant at a value $1/2$ for each spin state. Another special region in parameter space 
is the IN-EX boundary, characterized by the simultaneous presence of domain walls
in both spin states, as we discuss elsewhere \cite{reichenbach-deloc}.  

\begin{figure}
\begin{center}
\includegraphics[scale=1]{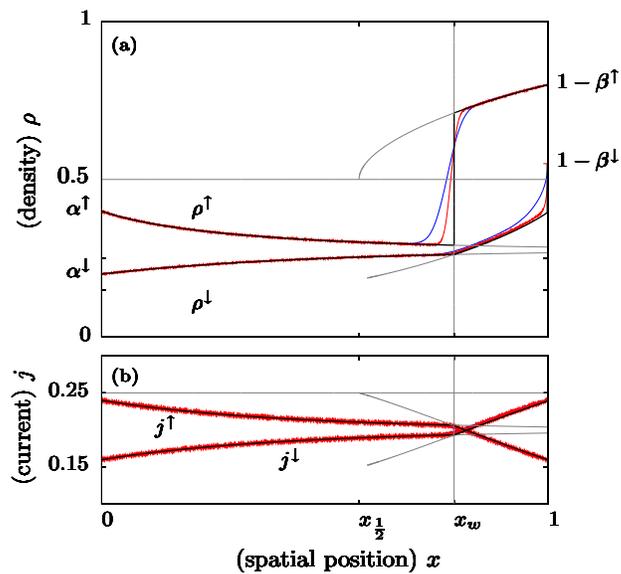}
\caption{(Color online) The densities (a) and currents (b) in the
IN region: generic state, exhibiting the polarization phenomenon.
Results from stochastic simulations are shown as blue ($L=2000$)
resp. red ($L=10000$) lines. They piecewise obey the first  order
approximations (black), grey lines indicate continuations of the
latter into regions where the densities are no longer given by
them. The parameters are
$\alpha^\uparrow=0.4,~\alpha^\downarrow=0.2,~\beta^\uparrow=0.2,~\beta^\downarrow=0.45$,
and $\Omega=0.5$.\label{EL}}
\end{center}
\end{figure}
Away from these regions, the generic situation for the density profiles is illustrated 
in~Fig. \ref{EL}. Here, we have considered parameters belonging to the IN region; the behavior in the EX region follows from particle-hole symmetry. A domain wall emerges for one spin state 
and a boundary layer for the other one. For specificity, we consider a domain wall for the spin-up state, the other situation is obtained from spin symmetry. The  density profiles $\rho_l^{\uparrow(\downarrow)}$
close to the left boundary are given by the solution of the first-order 
differential equations~(\ref{cont_a}),(\ref{cont_b}), obeying the left boundary conditions
 $\rho_l^\uparrow(x=0)=\alpha^\uparrow_\textnormal{\scriptsize eff}$ and 
$\rho_l^\downarrow(x=0)=\alpha^\downarrow_\textnormal{\scriptsize eff}$. 
For the density profiles $\rho_r^{\uparrow(\downarrow)}$ in the vicinity of the right boundary, we use the fact that  the current in bulk is determined by the injection rates, $J = J_\textnormal{\scriptsize IN}$ (which defines the IN
region). Therefore, the densities satisfy right boundary conditions which are given by 
 $\rho_r^\uparrow(x=1)=1-\beta^\uparrow_\textnormal{\scriptsize eff}$; and
$\rho_r^\downarrow(x=1)$ 
is found from the conservation of the
particle current:
\begin{eqnarray}\label{boundcond_b}
J&=&\alpha^\uparrow_{\textnormal{\scriptsize eff}}(1-\alpha^\uparrow_{\textnormal{\scriptsize eff}})
+\alpha^\downarrow_{\textnormal{\scriptsize eff}}(1-\alpha^\downarrow_{\textnormal{\scriptsize eff}})\cr
  &=&  \beta^\uparrow_{\textnormal{\scriptsize eff}} (1-\beta^\uparrow_{\textnormal{\scriptsize eff}})
+\rho^\downarrow_r(x=1)[1-\rho^\downarrow_r(x=1)] \, .
\end{eqnarray}

At some point $x_w$ in bulk, the left and right solutions have to be concatenated by a domain wall for spin-up.
To determine the position $x_w$ of this domain wall, we use the \emph{continuity} of the spin currents; see
Fig.~\ref{EL}(b). \footnote{Indeed, though they are not spatially conserved,
the mesoscopic scaling of the spin-flip rate $\omega$ was seen to
cause a only slowly varying spatial dependence; in the continuum
limit, the spin currents are continuous.} This continuity condition singles
out a distinct spatial position for the domain wall: Denote by
$\rho_l^\uparrow(x_w)$ the value of the density to the left of
$x_w$, and $\rho_r^\uparrow(x_w)$  the value to the right. From
$j^\uparrow=\rho^\uparrow(1-\rho^\uparrow)$ together with
$\rho^\uparrow_l(x_w)\neq\rho_r^\uparrow(x_w)$, we arrive at the
condition
\begin{equation}
\rho^\uparrow_l(x_w)=1-\rho_r^\uparrow(x_w)
\label{cond_xw}
\end{equation}
for the domain wall position. \footnote{For TASEP-like transport the
particle-hole symmetry restricts  the density jump to this mirror
relation. More general current-density relation are feasible
\cite{shaw-2003-68,pierobon-2006-74}, but are not expected to change the picture
qualitatively.} From the conservation of the particle current $J$, it  follows that the density $\rho^\downarrow$ is continuous at the position $x_w$.

When considering the internal states as actual spins, the
appearance of a domain wall in the density profile of one of the
spin states results in a \emph{spontaneous polarization
phenomenon}. Indeed, while both the density of spin-up and
spin-down remain at comparable low values in the vicinity of the
left boundary, this situation changes upon crossing the point
$x_w$. There, the density of spin-up jumps to a high value, while
the density of spin-down remains at a low value, resulting in a
polarization in this region.

Comparing the generic phase behavior to the one of TASEP, we
observe that the IN region can be seen as the analogue to the
low-density region there: within both, a low-density phase
accompanied by a boundary layer at the right boundary arises.
Following these lines, the EX region has its analogue in the
high-density region, while the MC region is straightforwardly
generalized from the one of TASEP. Furthermore, the delocalization
transition across the IN-EX boundary is similar to  the appearance
of a delocalized domain wall at the coexistence line in TASEP.

\subsection{Phases and phase boundaries}

\label{phases_bound}

In the generic situation of Fig.~\ref{EL},  the density of
spin-down is in a homogeneous low-density (LD) state, while for
spin-up, a low-density and a high-density region coexist. We refer
to the latter as the $\textnormal{LD-HD}_\textnormal{\scriptsize IN}$ phase, as the phase
separation arises within the IN region, to be contrasted  from a
$\textnormal{LD-HD}_\textnormal{\scriptsize EX}$ phase which may arise within the EX
region. Clearly, the $\textnormal{LD-HD}_\textnormal{\scriptsize IN}$ phase is only
present if the position $x_w$ of the domain wall lies within bulk.
Tuning the system's parameter, it may leave the system through the
left or right boundary, resulting in a homogeneous phase. Indeed,
$x_w=1$ marks the transition between the $\textnormal{LD-HD}_\textnormal{\scriptsize IN}$
phase and the pure LD state, while at $x_w=0$ the density changes
from the $\textnormal{LD-HD}_\textnormal{\scriptsize IN}$ to a homogeneous high-density
(HD) state. Regarding the domain wall position $x_w$ as an order
parameter, these transitions are continuous. Implicit analytic
expressions for these phase boundaries, derived in the following,
are obtained from the first-order approximation, Eqs.~(\ref{cont_a}) and (\ref{cont_b}).

Spin symmetry yields the analogous situation with a domain wall
appearing in the density profile of spin-down, while particle-hole
symmetry maps it to the EX region, where a pure HD phase arises
for one of the spins. Discontinuous transitions accompanied by
delocalized domain walls appear at the submanifold of the IN-EX
boundary (see~\cite{reichenbach-deloc} for a detailed
discussion).

The phase boundaries may be computed  from the condition $x_w=0$
and $x_w=1$ in the situation of Fig.~\ref{EL}. Consider first the
case of $x_w=0$. There, the density profiles are fully given
by the first-order approximation $\rho^{\uparrow(\downarrow)}_r$
satisfying the boundary conditions at the right. The condition
(\ref{cond_xw}) translates to
\begin{equation}
\rho^\uparrow_r(x=0)=1-\rho_l^\uparrow(x=0)=1-\alpha^\uparrow_{\textnormal{\scriptsize eff}}
\label{xw=0}
\end{equation}
which  yields an additional constraint on the system's parameters. This
defines the hyper-surface in the IN region where $x_w=0$ occurs,
and thus the phase boundary between the $\textnormal{LD-HD}_\textnormal{\scriptsize IN}$
and the pure HD phase.\\ Similarly, if $x_w=1$, the densities
follow the left solution $\rho^{\uparrow(\downarrow)}_l(x)$,
determined by the left boundary conditions, within the whole
system. From Eq. (\ref{cond_xw}) we obtain
\begin{equation}
\rho^\uparrow_l(x=1)=1-\rho_r^\uparrow(x=1)=\beta^\uparrow_{\textnormal{\scriptsize eff}}\, .
\label{xw=1}
\end{equation}
Again, the latter is a constraint on the parameters and defines
the hyper-surface in the IN region where $x_w=1$ is found, being
the phase boundary between the $\textnormal{LD-HD}_\textnormal{\scriptsize IN}$ and the
homogeneous LD phase.

The conditions (\ref{xw=0}), (\ref{xw=1}) yield implicit equations
for the phase boundaries. The phase diagram is thus determined up
to solving algebraic equations, which may be achieved numerically.
Further insight concerning the phase boundaries is possible and
may be obtained analytically, which we discuss next.

First, we note that in the case of equal injection rates, $\alpha^\uparrow=\alpha^\downarrow$,
the density profiles in the vicinity of the left boundary are constant. If in addition $\alpha^\uparrow=\alpha^\downarrow =\beta^\uparrow < 1/2$, we observe from Eq.~(\ref{xw=1}) that a domain wall at $x_w=1$ emerges. Therefore, this set of parameters always lies on the phase boundary $x_w=1$, independent of the value of $\Omega$.

Second, we investigate the phase boundary determined by $x_w=0$.
Comparing with Fig.~\ref{EL}, we observe that the first-order
approximation $\rho^\uparrow_r$ for the density of spin-up may
reach the value $\frac{1}{2}$ at a point which is denoted by
$x_{\frac{1}{2}}$: $\rho^\uparrow_r(x_{\frac{1}{2}})=\frac{1}{2}$.
This point corresponds to a branching point of the first-order
solution. Increasing $\Omega$, the value of $x_{\frac{1}{2}}$ increases as well. The domain wall in the
density of spin-up can only emerge at a value $x_w\geq
x_{\frac{1}{2}}$. At most, $x_w=x_{\frac{1}{2}}$, in which case a
domain wall with infinitesimal small height arises. For the
phase boundary specified by $x_w=0$, this implies that it only exists as long as
$x_{\frac{1}{2}}\leq 0$. The case  $x_w=x_{\frac{1}{2}}=0$
corresponds to a domain wall of infinitesimal height, which is
only feasible if $\alpha^\uparrow_{\textnormal{\scriptsize eff}}=\frac{1}{2}$. Now,
for given rates
$\alpha^\uparrow_{\textnormal{\scriptsize eff}}=\frac{1}{2},\alpha^\downarrow,\beta^\uparrow$,
the  condition $x_{\frac{1}{2}}=0$ yields a critical rate
$\Omega^*(\alpha^\downarrow,\beta^\uparrow)$, depending on the
rates $\alpha^\downarrow, \beta^\uparrow$. The situation $x_w=0$
can only emerge for rates
$\Omega\leq\Omega^*(\alpha^\downarrow,\beta^\uparrow)$. Varying
the rates $\alpha^\uparrow,\alpha^\downarrow$ and
$\beta^\uparrow$, the critical rate
$\Omega^*(\alpha^\downarrow,\beta^\uparrow)$ changes as well. In
\ref{app_first_order}, we show that its largest value occurs
at $\alpha^\downarrow=\beta^\uparrow=0$. They yield the rate
$\Omega_C\equiv\Omega^*(\alpha^\downarrow=\beta^\uparrow=0)$,
which is calculated to be
 \begin{equation}
\Omega_C=1+\frac{1}{4}\sqrt{2}\ln{(3-2\sqrt{2})}\approx 0.38  \, .
\end{equation}
The critical $\Omega^*(\alpha^\downarrow,\beta^\uparrow)$ are 
lying in the interval between $0$ and $\Omega_C$:
$\Omega^*(\alpha^\downarrow,\beta^\uparrow)\in [0,\Omega_C]$, and
all values in this interval in fact occur. The rate $\Omega_C$
defines a \emph{scale} in the spin-flip  rate $\Omega$: For
$\Omega\leq\Omega_C$, the phase boundary determined by $x_w=0$
exists, while disappearing  for $\Omega>\Omega_C$.

Third, we study the form of the phase boundaries for large
$\Omega$, meaning  $\Omega\gg\Omega_C$. In this case, the phase
boundary specified by $x_w=0$ is no longer present.
 Furthermore, it turns out that in this situation,  the
densities close to the left boundary quickly approximate a common value $\rho_e$. The latter
is found from conservation of the particle current:
$2\rho_e(1-\rho_e)=J$. We now consider  the implications for the
phase boundary determined by $x_w=1$. With
$\rho_l^\uparrow(x=1)=\rho_e$, Eq.~(\ref{xw=1}) turns into
$\rho_e=\beta^\uparrow_\textnormal{\scriptsize eff}$, yielding
\begin{equation}
2\beta^\uparrow_{\textnormal{\scriptsize eff}}(1-\beta^\uparrow_{\textnormal{\scriptsize eff}})=
\alpha^\uparrow_{\textnormal{\scriptsize eff}}(1-\alpha^\uparrow_{\textnormal{\scriptsize eff}})
+\alpha^\downarrow_{\textnormal{\scriptsize eff}}(1-\alpha^\downarrow_{\textnormal{\scriptsize eff}}) \, .
\label{xw1_largeomega}
\end{equation}
This condition specifies the phase boundary $x_w=1$, asymptotically for large $\Omega$. It constitutes a
simple quadratic equation in the in- and outgoing rates,
independent of $\beta^\downarrow$, and contains the set
$\alpha^\uparrow=\alpha^\downarrow_{\textnormal{\scriptsize eff}}=\beta^\uparrow$.

\section{Stochastic simulations}

\label{stoch_sim}

To confirm our analytic findings from the previous section, we
have performed stochastic simulations. The dynamical rules
(i)-(iv) described in Subsec.~\ref{dynamics} were implemented
using random sequential updating. In our simulations, we have
performed averages over typically $10^5$ time steps, with
$10\times L$ steps of updating between successive ones. Finite
size scaling singles out the analytic solution in the limit of
large system sizes, as exemplified in Figs.~\ref{eff_rates} and
\ref{EL}.

For all  simulations, we have checked that the analytic
predictions are recovered upon approaching the mesoscopic limit.
We attribute the apparent exactness of our analytic approach in
part to the exact current density relation in the steady state of
the TASEP~\cite{derrida-1998-301}. The additional coupling of the two
TASEPs in our model is only weak: the local exchange between the
two states vanishes in the limit of large system sizes.
Correlations between them are washed out, and mean-field is
recovered.

The observed exactness of the analytic density profiles within the
mesoscopic limit implies that our analytic approach yields exact
phase diagrams as well. The latter are the subject of the
subsequent section.

\section{Two-dimensional phase diagrams}

\label{phase_diag}

In this section, we discuss the phase behavior on two-dimensional
cuts in the whole five-dimensional parameter space. Already  the
simplified situation of equal injection  rates,
$\alpha^\uparrow=\alpha^\downarrow$, yields interesting behavior.
There as well as in the general case, we investigate the role of
the spin-flip rate $\Omega$ by discussing the situation of small
and large values of $\Omega$.

\subsection{Equal injection rates}

For simplicity, we start our discussion of the phase diagram with equal injection rates, $\alpha^\uparrow=\alpha^\downarrow$. Then, the spin polarization phenomenon, depicted in Fig.~\ref{EL},
becomes even more striking. Starting from equal densities at the left boundary, and hence zero polarization, spin polarization suddenly switches on at the domain wall position $x_w$.
The particular location of $x_w$ is not triggered by a cue on the track, but tuned through the model parameters.

The phase transitions from LD to the
$\textnormal{LD-HD}_\textnormal{\scriptsize IN}$ arising in the IN region take a
remarkably simple form. Their location is found from $x_w=1$, and is determined by
Eq.~(\ref{xw=1}) (if phase coexistence arises for spin-up). Since $\rho^\uparrow(x) = \rho^\downarrow(x) =
\alpha = \textnormal{const.}$ for $x < x_\textnormal{\scriptsize w}$, Eq.~(\ref{xw=1}) turns into
 $\alpha = \beta^\uparrow$. The latter transition line
intersects the IN-EX boundary, given by $J_\textnormal{\scriptsize IN}=J_\textnormal{\scriptsize EX}$,
at $\beta^\uparrow=\beta^\downarrow=\alpha$, i.e. at the point
where all entrance and exit rates coincide. At this
\emph{multicritical point} $\mathcal{A}$, a continuous line
intersects a discontinuous one. The same transition in the density
of spin-down state is, from similar arguments, located at $\alpha
= \beta^\downarrow$, and also coincides with the IN-EX boundary in
$\mathcal{A}$. Neither the multicritical point $\mathcal{A}$ nor
these phase boundaries depend on the magnitude of the gross spin
flip rate $\Omega$. Therefore, qualitatively tuning the system's state
is possible only upon changing the injection or extraction rates.
The other phase transitions  within the IN region, namely from the
HD to the $\textnormal{LD-HD}_\textnormal{\scriptsize IN}$ phase, are more involved. The
analytic solution (\ref{x_rhoa}), (\ref{x_rhob}) has to be
considered together with the condition (\ref{xw=0}) for the
transition. However, at the end of Subsec. \ref{phases_bound}, we have found that
these transitions (determined by $x_w=0$)  disappear for sufficiently
large~$\Omega>\Omega_C$.

\subsubsection{Large values of $\Omega$.}

\begin{figure}
\begin{center}
\includegraphics[scale=1]{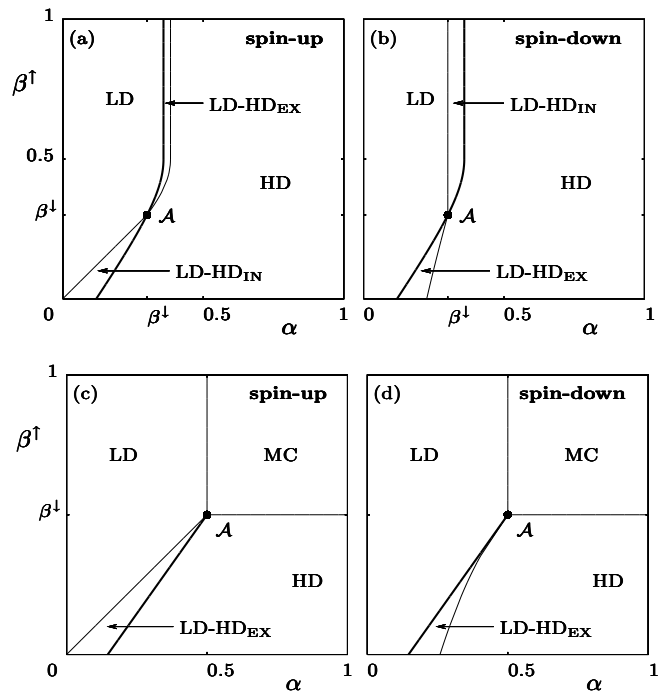}
\caption{ Phase diagrams in the situation of equal entrance rates
$\alpha^\uparrow=\alpha^\downarrow\equiv\alpha$ and large
$\Omega$. The phases of the densities of spin-up (spin-down) state
are shown in (a) resp. (b) for a value $\beta^\downarrow=0.3$. At
a multicritical point $\mathcal{A}$, continuous lines (thin)
intersect with a discontinuous line (bold), the IN-EX boundary. If
$\beta^\downarrow\geq\frac{1}{2}$, the maximal current phase
appears for spin-up, see (c), as well for spin-down, drawn in (d).
In the first situation, the switching rate is $\Omega=0.15$, while
$\Omega=0.2$ in the second.\label{phasediag_eqal_largeomega}}
\end{center}
\end{figure}
In the situation of large $\Omega>\Omega_C$, phase transitions
arising from $x_w=0$ in the IN region or from the analogue in the
EX region do not emerge, as discussed at the end of Subsec.~\ref{phases_bound}.
We have drawn resulting phase diagrams in
Fig.~\ref{phasediag_eqal_largeomega}, showing the phase of spin-up
(spin-down) in the left (right) panels, depending on $\alpha$ and
$\beta^\uparrow$. Along the IN-EX boundary, being the same line
(shown as bold) in the left and right panels, a delocalization
transitions occur. At the multicritical point $\mathcal{A}$, it is
intersected by continuous lines emerging within the IN resp. the
EX region. When $\beta^\downarrow>1/2$, a maximal current(MC)
phase emerges in the upper right quadrant, see
Fig.~\ref{phasediag_eqal_largeomega} (c)-(d).

To illustrate the system's phase behavior, let us consider what happens along a horizontal line in the phase diagrams (a) and (b), at a value  $\beta^\uparrow>\beta^\downarrow$. At such a line, for small values of $\alpha$, both spin states are in low-density (LD) phases. Upon crossing a certain value of $\alpha$, a domain wall enters at $x_w=1$ in the spin-down density profile. Then, spin-down exhibits phase coexistence ($\textnormal{LD-HD}_\textnormal{\scriptsize IN}$), while spin-up remains in a LD phase. Further increasing $\alpha$, the bold line is reached, where delocalized domain walls arise in both spin states. For larger values of $\alpha$, a localized domain wall emerges for spin-up (implying a $\textnormal{LD-HD}_\textnormal{\scriptsize EX}$ phase), and a pure HD phase for spin-down. If $\alpha$ is further increased, the domain wall in the spin-up density profile leaves the system through the left boundary (at $x_w=0$), and pure HD phases remain for both spin states.

While we have found the transitions within
the IN region by simple expressions in the previous subsection, the ones emerging in the EX
region are more complex and involve the full analytic solution
(\ref{x_rhoa}), (\ref{x_rhob}). Their most notable feature is that
the width of the corresponding coexistence phase decreases with increasing spin-flip rate $\Omega$, until it finally vanishes in the limit $\Omega
\to \infty$. This may be seen by considering the analogue of
Eq.~(\ref{xw1_largeomega}) in the EX region, which describes the
phase boundary as it is asymptotically approached when
$\Omega\rightarrow\infty$:
\begin{equation}
2\alpha_{\textnormal{\scriptsize eff}}(1-\alpha_{\textnormal{\scriptsize eff}})=\beta^\uparrow_{\textnormal{\scriptsize eff}}
(1-\beta^\uparrow_{\textnormal{\scriptsize eff}})+\beta^\downarrow_{\textnormal{\scriptsize eff}}(1-\beta^\downarrow_{\textnormal{\scriptsize eff}})
\, ;
\end{equation}
it coincides with the IN-EX boundary.

\subsubsection{Small values of $\Omega$.}

\begin{figure}
\begin{center}
\includegraphics[scale=1]{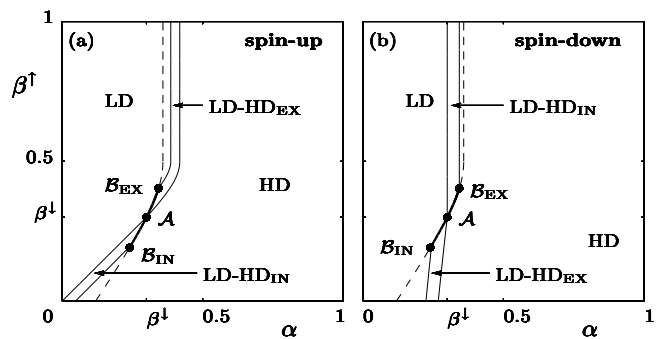}
\caption{Phase diagrams in the situation of
Fig.~\ref{phasediag_eqal_largeomega} (a), (b), but with $\Omega$
decreased to a small value, $\Omega=0.05$. Additional phase
transitions emerge in the IN as well as the EX region, accompanied
by multicritical points $\mathcal{B}_\textnormal{\scriptsize IN}$ and
$\mathcal{B}_\textnormal{\scriptsize EX}$. Caused by them,  phase transitions do no
longer appear  across some parts of the IN-EX boundary, which is
there shown as dashed line. \label{phasediag_eqal_smallomega}}
\end{center}
\end{figure}
As discussed at the end of Subsec.~\ref{phases_bound}, when $\Omega<\Omega_C$, the appearance of additional phase
transitions becomes possible. For example, within the IN region, the situation
$x_w=0$ may emerge. It describes the transition from the HD to the
$\textnormal{LD-HD}_\textnormal{\scriptsize IN}$  phase; the analogue occurs in the EX
region. In Fig.~\ref{phasediag_eqal_smallomega}, we show resulting phase diagrams for the spin-up (left panel) and spin-down (right panel), resp.. The additional transition lines 
intersect the IN-EX boundary (bold) at additional multicritical
points $\mathcal{B}_\textnormal{\scriptsize IN}$ and $\mathcal{B}_\textnormal{\scriptsize EX}$. Also,
they partly substitute the IN-EX boundary as a phase boundary:
across some parts of the latter, phase transitions do not arise.
This behavior reflects the \emph{decoupling} of the two states for
decreasing spin-flip rate $\Omega$. Indeed, for $\Omega\rightarrow
0$, the states become more and more decoupled, such that the IN-EX
boundary, involving the combined entrance and exit rates of both
states, loses its significance.

\subsubsection{Multicritical points.}

Although the shapes of most of the transition lines appearing in
the phase diagrams shown in Fig.~\ref{phasediag_eqal_smallomega}
are quite involved, they also exhibit simple behavior.
\emph{Pairwise}, namely one line from a transition in spin-up and
another from a related transition in spin-down states, they
intersect the IN-EX boundary in the same multicritical point. This
intriguing phenomenon may be understood by considering the
multicritical points: e.g., at $\mathcal{A}$, the transition line
from the LD to the $\textnormal{LD-HD}_\textnormal{\scriptsize IN}$  phase in the density
profile of spin-up intersects the IN-EX boundary, which implies
that there we have a domain wall in the density profile of spin-up
at the position $x_w=1$. However, being on the IN-EX boundary, the condition $J_\textnormal{\scriptsize IN}=J_\textnormal{\scriptsize EX}$ implies that in this situation a domain
wall forms as well in the density of spin-down states, also
located at $x_w=1$. Consequently,  $\mathcal{A}$ also marks the point
where the transition line specified by $x_w=1$ for spin-down
states intersects the IN-EX boundary. Due to the special situation
of equal entrance rates, one more pair of lines intersects in this
point.  Similarly, at $\mathcal{B}_\textnormal{\scriptsize IN}$, the transition line
from the HD to $\textnormal{LD-HD}_\textnormal{\scriptsize IN}$  phase in the density
profile of spin-up intersects the IN-EX boundary, such that a
domain wall forms in the density of spin-up at $x_w=0$. Again, as
$J_\textnormal{\scriptsize IN}=J_\textnormal{\scriptsize EX}$ holds on the IN-EX boundary, this implies
the formation of a domain wall in the density of spin-down at
$x_w=1$, corresponding to the transition from the LD to the
$\textnormal{LD-HD}_\textnormal{\scriptsize EX}$ phase for spin-down in the EX region.

\subsection{The general case}

Having focused on the physically particularly enlightening case of
equal entering rates in the previous subsection, we now turn to
the general case. To illustrate our findings, we show phase
diagrams depending on the injection and extraction rates for
spin-up states, $\alpha^\uparrow$ and $\beta^\uparrow$. Similar
behavior as for equal entrance rates is observed. The
multicritical point $\mathcal{A}$ now splits up into two distinct
points $\mathcal{A}_\textnormal{\scriptsize IN}$ and $\mathcal{A}_\textnormal{\scriptsize EX}$.

\subsubsection{Large values of $\Omega$: Asymptotic results.}
\label{large_omega}

\begin{figure}
\begin{center}
\includegraphics[scale=1]{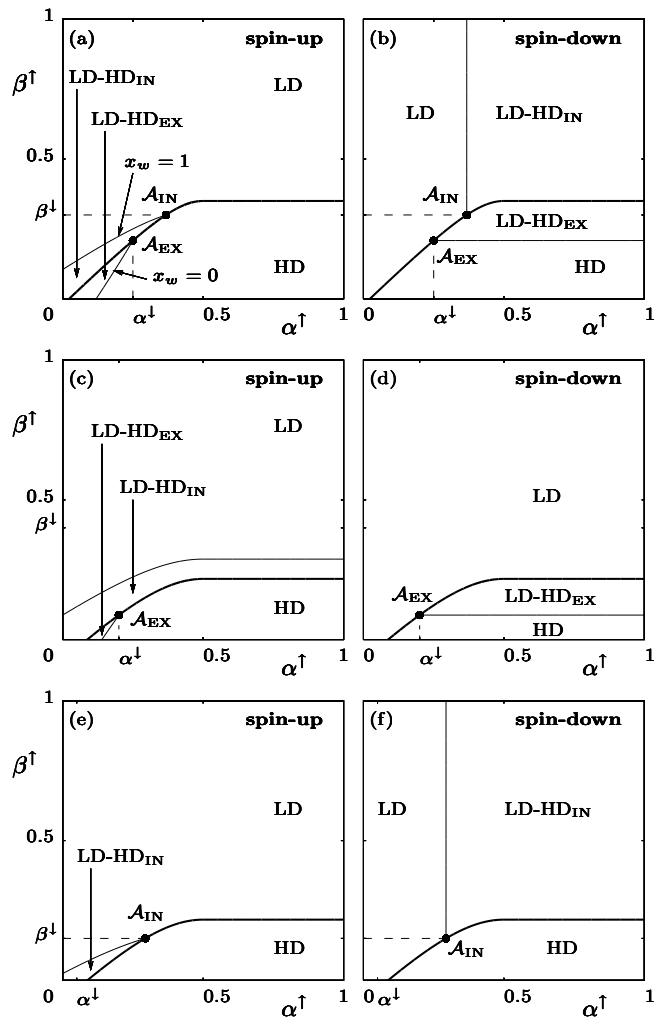}
\caption{Phase diagrams in the general situation: asymptotic
results for large $\Omega$. Lines of continuous transitions (thin)
within the IN resp. EX region intersect the delocalization
transition line (bold) in multicritical points
$\mathcal{A}_\textnormal{\scriptsize IN}$ resp.  $\mathcal{A}_\textnormal{\scriptsize EX}$. Both of
these points appear in (a), (b)
($\alpha^\downarrow=0.25,\beta^\downarrow=0.3$) while only
$\mathcal{A}_\textnormal{\scriptsize EX}$ is present in (c), (d)
($\alpha^\downarrow=0.2,\beta^\downarrow=0.4$) and
$\mathcal{A}_\textnormal{\scriptsize IN}$ alone in (e), (f)
($\alpha^\downarrow=0.05,\beta^\downarrow=0.15$), yielding
different topologies.   \label{phasediag_largeomega1}}
\end{center}
\end{figure}
\begin{figure}
\begin{center}
\includegraphics[scale=1]{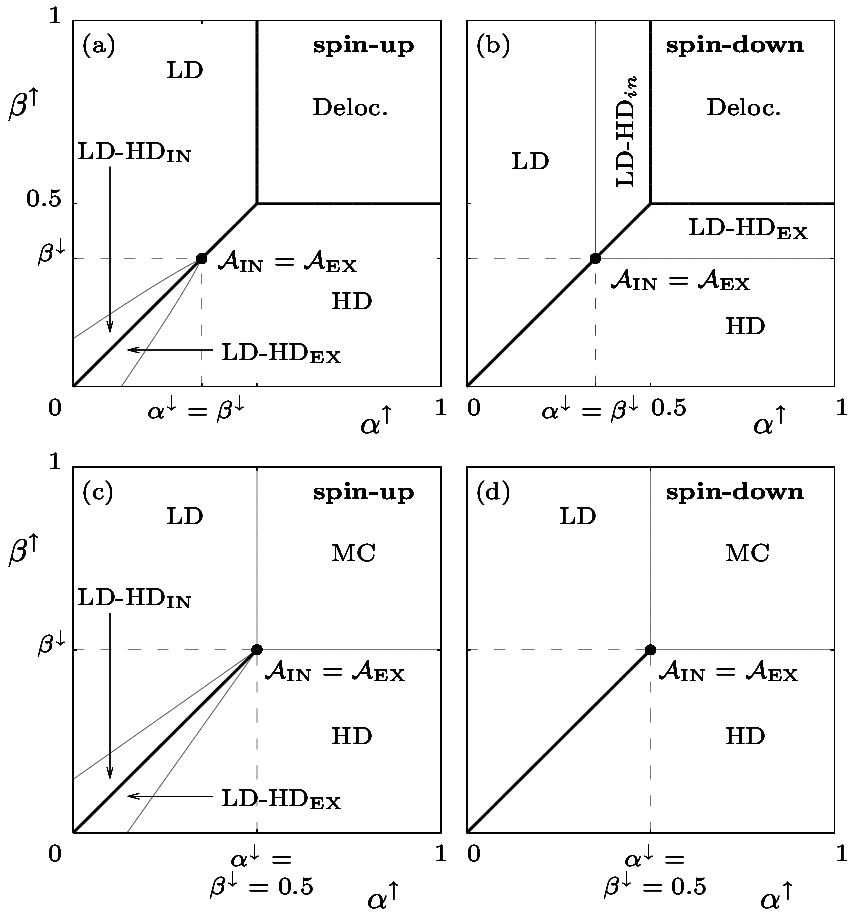}
\caption{Delocalization as well as maximal current phase (MC).
When $\alpha^\downarrow=\beta^\downarrow<1/2$ and
$\alpha^\uparrow,\beta^\uparrow\geq 1/2$ [upper right quadrant in
(a), (b)], delocalized domain walls form in the density profiles
of both spin states. If instead
$\alpha^\downarrow,\beta^\downarrow\geq 1/2$, the maximal current
phase emerges, see (c), (d).\label{phasediag_largeomega2}}
\end{center}
\end{figure}
Again, large $\Omega$ prohibit the emergence of the phase
transition from the HD to the $\textnormal{LD-HD}_\textnormal{\scriptsize IN}$  phase in
the IN region as well as from the LD to the
$\textnormal{LD-HD}_\textnormal{\scriptsize EX}$ phase within the EX region, see end of Subsec.~\ref{phases_bound}. In this
paragraph, we consider phase diagrams which are approached
asymptotically when $\Omega\rightarrow\infty$. Convergence is fast
in $\Omega$, and the asymptotic phase boundaries yield an
excellent approximation already for $\Omega\gtrsim 2\Omega_C$.

The transition from LD to the $\textnormal{LD-HD}_\textnormal{\scriptsize IN}$  phase in
the IN region asymptotically takes the form of
Eq.~(\ref{xw1_largeomega}), and the one from HD to the
$\textnormal{LD-HD}_\textnormal{\scriptsize EX}$  phase in the EX region is obtained by
particle-hole symmetry. All phase boundaries, including the IN-EX
boundary, are thus given by simple quadratic
expressions.

Phase diagrams with different topologies that can emerge are
exhibited in Fig.~\ref{phasediag_largeomega1} and
\ref{phasediag_largeomega2}. As in the previous subsection, we
show the phases of spin-up (spin-down) states on the left (right)
panels. The phase boundaries between the LD and the
$\textnormal{LD-HD}_\textnormal{\scriptsize IN}$ phase in the IN region for spin-up as
well as spin-down both intersect the IN-EX boundary in a
multicritical point $\mathcal{A}_\textnormal{\scriptsize IN}$, being located at
$\beta^\uparrow=\beta^\downarrow$. Similarly, the lines of
continuous transitions within the EX region both coincide with the
IN-EX boundary in a multicritical point $\mathcal{A}_\textnormal{\scriptsize EX}$,
which is situated at $\alpha^\uparrow=\alpha^\downarrow$. Note
that the phase transitions emerging in the density profile of
spin-down within the IN region do not depend on $\beta^\uparrow$,
thus being horizontal lines in the phase diagrams. Within the
EX region they are independent of $\alpha^\uparrow$, yielding
vertical lines.

For $\alpha^\downarrow, \beta^\downarrow<1/2$,
Fig.~\ref{phasediag_largeomega1} shows different  topologies of
phase diagrams, which only depend on which of the multicritical
points   $\mathcal{A}_\textnormal{\scriptsize IN}$, $\mathcal{A}_\textnormal{\scriptsize EX}$ is
present.  If both appear, see
Fig.~\ref{phasediag_largeomega1} (a), (b), the
$\textnormal{LD-HD}_\textnormal{\scriptsize IN}$ and the $\textnormal{LD-HD}_\textnormal{\scriptsize EX}$ phase
for spin-up are adjacent to each other, separated by the IN-EX
boundary. Although in both phases localized domain walls emerge,
their position changes discontinuously upon crossing the
delocalization transition. E.g., starting within the
$\textnormal{LD-HD}_\textnormal{\scriptsize IN}$ phase,  the domain wall delocalizes when
approaching the IN-EX boundary,  and, having crossed it,
relocalizes again, but at a different position.

When $\alpha^\downarrow=\beta^\downarrow<1/2$, a subtlety emerges,
see Fig.~\ref{phasediag_largeomega2} (a), (b). If both
$\alpha^\uparrow\geq 1/2$ and $\beta^\uparrow\geq 1/2$, i.e. in
the upper right quadrant of the phase diagrams, these rates
effectively act as $1/2$, and the condition $J_\textnormal{\scriptsize IN}=J_\textnormal{\scriptsize EX}$ for the
IN-EX boundary is fulfilled \emph{in this whole region.}
Therefore, delocalized domain walls form on both lanes within this
region, as is confirmed by our stochastic
simulations~\cite{reichenbach-deloc}.

The maximal current phase (MC) emerges when all rates exceed or
equal the value $1/2$, corresponding to the upper left quadrant of
the phase diagrams in Fig.~\ref{phasediag_largeomega2} (c), (d).

\subsubsection{Small values of $\Omega$.}

\begin{figure}
\begin{center}
\includegraphics[scale=1]{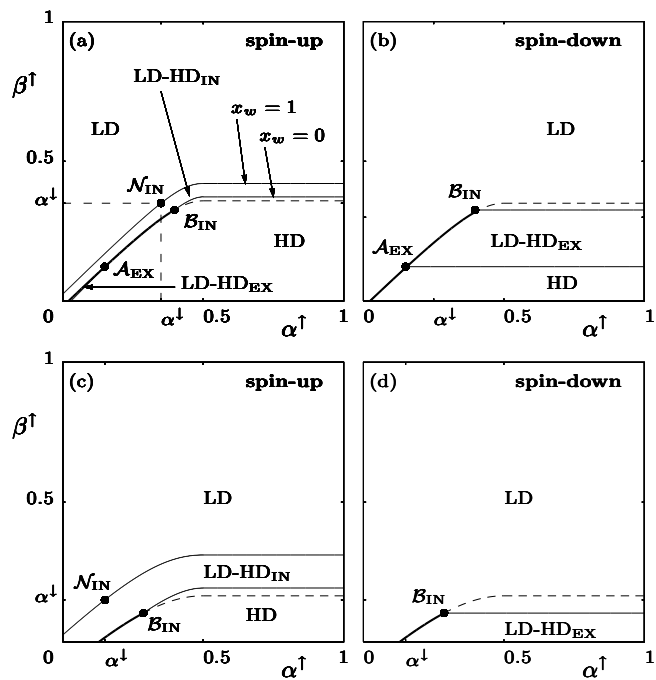}
\caption{Phase diagrams in the general case and small values of
$\Omega$. The nodal point $\mathcal{N}_\textnormal{\scriptsize IN}$ remains
unchanged when $\Omega$ is varied. The appearance of the
multicritical point $\mathcal{B}_\textnormal{\scriptsize IN}$ is accompanied by the
non-occurrence of phase transitions across parts of the IN-EX
boundary, then shown as dashed line. The multicritical point
$\mathcal{A}_\textnormal{\scriptsize EX}$ emerges in (a), (b), but not in the
situation of (c)(, (d). Parameters are
$\Omega=0.08,~\alpha^\downarrow=0.35,~\beta^\downarrow=0.45$ in
(a), (b) and
$\Omega=0.2,~\alpha^\downarrow=0.15,~\beta^\downarrow=0.4$ in (c),
(d). \label{phasediag_smallomega}}
\end{center}
\end{figure}
When $\Omega<\Omega_C$, the transitions from LD to
$\textnormal{LD-HD}_\textnormal{\scriptsize IN}$ within the IN region as well as the
analogue in the EX region are possible. As in the case of equal
entering rates, the corresponding transition lines pairwise
intersect the IN-EX boundary in multicritical points
$\mathcal{B}_\textnormal{\scriptsize IN}$ and $\mathcal{B}_\textnormal{\scriptsize EX}$. As all
transitions between phases of the spin-down density  within the IN region are
independent of $\beta^\uparrow$, the corresponding lines are
simply horizontal; and within the EX region, their independence of
$\alpha^\uparrow$ implies that they yield vertical lines. The
phase diagram for the density of spin-down is thus easily found
from the IN-EX boundary given by $J_\textnormal{\scriptsize IN}=J_\textnormal{\scriptsize EX}$ together
with the locations of the multicritical points
$\mathcal{A}_\textnormal{\scriptsize IN}$, $\mathcal{A}_\textnormal{\scriptsize EX}$,
$\mathcal{B}_\textnormal{\scriptsize IN}$ and $\mathcal{B}_\textnormal{\scriptsize EX}$. The latter
follow from the intersection of phase transition lines for the
density of spin-up, involving the whole analytic solution
(\ref{x_rhoa}), (\ref{x_rhob}), with the IN-EX boundary.

In Fig.~\ref{phasediag_smallomega} two interesting topologies that
may arise are exemplified. Induced by the presence of the
multicritical point $\mathcal{B}_\textnormal{\scriptsize IN}$, phase transitions do
not occur across all the IN-EX boundary, which is then only shown as
dashed line.  In Fig.~\ref{phasediag_smallomega} (a), (b),  the
points $\mathcal{A}_\textnormal{\scriptsize EX}$ and $\mathcal{B}_\textnormal{\scriptsize IN}$ are
present. The $\textnormal{LD-HD}_\textnormal{\scriptsize IN}$ phase for spin-up intervenes
the LD and the HD phase;  the $\textnormal{LD-HD}_\textnormal{\scriptsize EX}$ phase for
spin-up is also present, though very tiny. In the phase diagram of
spin-down, the $\textnormal{LD-HD}_\textnormal{\scriptsize IN}$ phase intervenes the LD
and the HD phase accompanied by continuous as well as
discontinuous transitions. Again, the presence of the
multicritical points induces the topology; e.g. in
Fig.~\ref{phasediag_smallomega} (c), (d), only
$\mathcal{B}_\textnormal{\scriptsize IN}$ appears. For the discussion of the
possible topologies, we encounter the restriction that
$\mathcal{A}_\textnormal{\scriptsize IN}$ and $\mathcal{B}_\textnormal{\scriptsize IN}$ cannot occur
together, as well as $\mathcal{A}_\textnormal{\scriptsize EX}$ and
$\mathcal{B}_\textnormal{\scriptsize EX}$ exclude each other (otherwise, the lines
determined by $x_w=0$ and $x_w=1$ would cross). 

We now discuss the  influence of the spin-flip  rate $\Omega$ on
the continuous transition lines for spin-up. In
Subsec.~\ref{phases_bound} the manifold defined by
$\alpha^\uparrow=\beta^\uparrow=\alpha^\downarrow_{\textnormal{\scriptsize eff}}$
was found to be a sub-manifold of the phase boundary specified by
$x_w=1$ in the IN region.  Independent of $\Omega$, the point
$\alpha^\uparrow=\beta^\uparrow=\alpha^\downarrow_{\textnormal{\scriptsize eff}}$,
denoted by $\mathcal{N}_\textnormal{\scriptsize IN}$, thus lies on the boundary
between the LD and the $\textnormal{LD-HD}_\textnormal{\scriptsize IN}$ phase (determined by $x_w=1$).  For large
$\Omega$, this boundary approaches the one given by
Eq.~(\ref{xw1_largeomega}).\\ Regarding the transition from the HD
to the $\textnormal{LD-HD}_\textnormal{\scriptsize IN}$  within the IN region (determined by $x_w=0$),
Subsec.~\ref{phases_bound} revealed that for increasing $\Omega$
it leaves the IN region at a critical transfer rate
$\Omega^*(\alpha^\downarrow, \beta^\uparrow)$. In the limit $\Omega\rightarrow 0$, the
densities $\rho^\uparrow(x)$ and $\rho^\downarrow(x)$ approach
constant values, and both the curve $x_w=1$ as $x_w=0$ for spin-up
in the IN region approach the line
$\beta^\uparrow=\alpha^\uparrow$ for $\alpha^\uparrow\leq
\frac{1}{2}$;  The phase in the upper right quadrant in the phase
diagram converges to the MC phase, such that in this limit,  the
case of two uncoupled TASEPs is recovered.

\section{Conclusions}
 
\label{concl}

We have presented a detailed study of an exclusion process with
internal states recently introduced in~\cite{reichenbach-2006-97}.
The Totally Asymmetric Exclusion Process (TASEP) has been
generalized by assigning  two internal states to the particles.
Pauli's exclusion principle allows double occupation  only for
particles in different internal states. Occasional switches from
one internal state to the other induce a coupling between the
transport processes of the separate states. Such a  dynamics
encompasses diverse situations, ranging from  vehicular traffic on
multiple lanes to molecular motors walking on intracellular tracks
and future spintronics devices.

We have elaborated on the properties of the emerging
non-equilibrium steady state focusing on density and current
profiles. In a mesoscopic scaling of the switching rate between
the internal states, nontrivial phenomena emerge. A localized
domain wall in the density profile of one of the internal states
induces a spontaneous polarization effect when viewing the
internal states as spins. We provide an explanation based on  the
weakly conserved currents of the individual states and the
current-density relations. A quantitative analytic description
within  a mean-field approximation and a continuum limit has been
developed and solutions for the density and current profiles have
been presented. A comparison with stochastic simulations revealed
that our analytic approach becomes exact in the limit of large
system sizes. We have attributed this remarkable finding to the
exact current-density relation in the TASEP, supplemented by the
locally weak coupling of the two TASEPs appearing in our model:
$\omega\rightarrow 0$ in the limit of large system sizes. Local
correlations between the two internal states are thus obliterated,
as particles hop forward on  a much faster timescale than they
switch their internal state.

Furthermore, the parameter regions that allow for the formation of
a localized domain wall have been considered. Analytic phase
diagrams for various scenarios, in particular the case of equal
entrance rates, have been derived. The phase diagrams have been
found to exhibit a rich structure, with continuous as well as
discontinuous non-equilibrium phase transitions. The discontinuous
one originates in the conserved particle current, which is either
limited by injection or extraction of particles. At the
discontinuous transition between both regimes, delocalized domain
walls emerge in the density profiles of \emph{both} internal
states. Multicritical points appear at the intersections of
different transition lines organizing the topology of the phase
diagrams. Two classes of multicritical points  are identified, one
of them  arises only  for sufficiently small gross spin-flip rate
$\Omega<\Omega_C$. The value $\Omega_C$, calculated analytically,
provides a natural scale for the rate $\Omega$.

It would be of interest to see which of the described phenomena
qualitatively remain when generalizing the model to include more
than two internal states. Indeed,  within the context of molecular
motors walking on microtubuli~\cite{Howard},  between 12 and 14
parallel lanes are relevant. Also, the internal states might
differ in the sense of different switching rates from one to
another~\cite{pronina-2006-372}  and the  built-in asymmetry may
result in different phases. In the context of intracellular
transport it appears worthwhile to investigate  the consequences
of a coupling to a bulk reservoir, c.f.
\cite{klumpp-2003-113,parmeggiani-2003-90,parmeggiani-2004-70}; in particular,
to study the interplay of   domain wall formation induced by  attachment and
detachment processes as well as rare switching events.

\section{Acknowledgements}

The authors are grateful for helpful discussions with Felix
von Oppen, Ulrich Schollw\"ock, Paolo Pierobon and Mauro Mobilia.
Financial support of the German Excellence Initiative via the program ``Nanosystems Initiative Munich (NIM)'' is gratefully acknowledged.

\appendix

\section{The densities in the first order approximation and
the critical value $\Omega_C$\label{app_first_order}}

In this Appendix, we give details on the derivation of the
analytic solution of the mean-field approximation in the continuum
limit to
 first order in $\epsilon$, i.e.  the system of differential equations (\ref{cont_a}),
  (\ref{cont_b}).

Summing them we find
\begin{eqnarray}
\partial_x[2\rho^\uparrow(x_i)-1]^2 +\partial_x[2\rho^\downarrow(x_i)-1]^2=0\, , \label{start}
\end{eqnarray}
such that
\begin{equation}
[2\rho^\uparrow(x_i)-1]^2 +[2\rho^\downarrow(x_i)-1]^2=J\, ,
\end{equation}
constitutes a first integral. Remember that
$j^\textnormal{\scriptsize tot}=\rho^\uparrow(x_i)[1-\rho^\uparrow(x_i)]+\rho^\downarrow(x_i)[1-\rho^\downarrow(x_i)]$,
such that  $J$  is given by the total current:
\begin{equation}
J=2-4j^\textnormal{\scriptsize tot} \, .
\end{equation}
This equation suggests the following parameterization:
\begin{eqnarray}
\cos{\theta}=&J^{-\frac{1}{2}}(2\rho^\uparrow-1) \cr
\sin{\theta}=&J^{-\frac{1}{2}}(2\rho^\downarrow-1)  \, .
\end{eqnarray}
The derivative reads
\begin{equation}
\frac{\sqrt{J}}{2}\cos{\theta}\frac{d\theta}{dx}=\frac{d\rho^\uparrow}{dx}
\, ,
\end{equation}
which leads to the differential equation
\begin{equation}
\sqrt{J}\sin{\theta}\cos{\theta}\frac{d\theta}{dx}=\Omega(\sin\theta-\cos\theta)
\, .
\end{equation}
This may be solved by a separation of  variables:
\begin{equation}
\frac{\Omega}{\sqrt{J}}x=\int_{\theta(0)}^{\theta(x)}\frac{\sin\theta\cos\theta}{\sin\theta-\cos\theta}d\theta
\, . \nonumber
\end{equation}
To perform the integral, the substitution $y=\tan\frac{\theta}{2}$
is useful. We obtain the inverse function $x=x(\theta)$:
\begin{equation}
x(\theta)=\frac{\sqrt{J}}{\Omega}G(y)\bigg|_{y=\tan{\frac{\theta}{2}}}+I\,
. \label{x_theta}
\end{equation}
Here we defined the function $G(y)$ by 
\begin{eqnarray}
G(y)=\bigg\{\frac{1+y}{1+y^2}
+\frac{\sqrt{2}}{4}\ln{\bigg|\frac{\sqrt{2}-(1+y)}{\sqrt{2}+1+y}\bigg|}\bigg\}\,
, \label{G}
\end{eqnarray}
and $I$ is an constant of integration.

To obtain the inverse functions
$x(\rho^\uparrow)$ and $x(\rho^\downarrow)$, we have to express
$\tan\frac{\theta}{2}$ by $\rho^\uparrow$ resp. $\rho^\downarrow$. Recognize
that $\tan\frac{\theta}{2}$ can be positive or negative. We therefore define
\begin{equation}
s^\downarrow=\left\{\begin{array}{cl}
-1 & \textnormal{if}~ \rho^\downarrow <\frac{1}{2} \\
+1 & \textnormal{if}~ \rho^\downarrow >\frac{1}{2}
\end{array}  \right. \ \, ,
\end{equation}
and analogously $s^\uparrow$ with $\uparrow$ and $\downarrow$ interchanged. Now
\begin{equation}
\tan\frac{\theta}{2}=s^\downarrow\cdot
\sqrt{\frac{1-J^{-1/2}(2\rho^\uparrow-1)}{1+J^{-1/2}(2\rho^\uparrow-1)}}
\, .
\end{equation}
The inverse functions $x(\rho^\uparrow)$ and $x(\rho^\downarrow)$ thus read:
 \begin{eqnarray}
x(\rho^\uparrow)=&\frac{\sqrt{J}}{\Omega}G(y)
\Big|_{y=s^\downarrow\cdot\sqrt{\frac{1-J^{-1/2}(2\rho^\uparrow-1)}{1+J^{-1/2}(2\rho^\uparrow-1)}}}+I
\label{x_rhoa}    \\
x(\rho^\downarrow)=&\frac{\sqrt{J}}{\Omega}G(y)
\Big|_{y=s^\uparrow\cdot\sqrt{\frac{1-J^{-1/2}(2\rho^\downarrow-1)}{1+J^{-1/2}(2\rho^\downarrow-1)}}}+I
\, . \label{x_rhob}
\end{eqnarray}
The  constants of integration $I$ and $J=2-4j^\textnormal{\scriptsize tot}$ are
determined by matching the boundary conditions. The inverse
functions of Eqs.~(\ref{x_rhoa}), (\ref{x_rhob}) constitute the
solution to Eqs.~(\ref{cont_a}) and (\ref{cont_b}), within the
first-order approximation to the mean-field equations for the
densities in the continuum limit.\\

Next, we  derive the result on $\Omega_C$ given at the end of
Subsec.~\ref{phases_bound}. Therefore, consider Fig.~\ref{EL}. We
are interested in the point $x_{\frac{1}{2}}$, and thus in the
right branch of the spin-up density profile. As the spin-down
density is in the LD phase, i.e. it is smaller than $\frac{1}{2}$,
we have $s^\downarrow=-1$ in the above solution for
$\rho^\uparrow$. Thus, $y\leq 0$ in Eq.~(\ref{x_rhoa}). At the
branching point of the analytic solution, i.e. the point
$x_{\frac{1}{2}}$, we have the density $\frac{1}{2}$, such that
there $y=-1$, implying $G(y)=0$. Now, if this branching point lies
on the right boundary, $x_{\frac{1}{2}}=0$, as it does for the
critical $\Omega^*$, this yields $I=0$ in Eq.~(\ref{x_rhoa}). On
the other hand, the right branch satisfies the boundary condition
on the right: $\rho^\uparrow(x=1)=1-\beta^\uparrow$. Upon
substitution into Eq.~(\ref{x_rhoa}), we obtain
\begin{equation}
1=x(1-\beta^\uparrow)=\frac{\sqrt{J}}{\Omega^*}G(y)
\Big|_{y=-\sqrt{\frac{1-J^{-1/2}(1-2\beta^\uparrow)}{1+J^{-1/2}(1-2\beta^\uparrow)}}}
\, .
\end{equation}  
which for given $\alpha^\downarrow,~ \beta^\uparrow$ is an
equation for $\Omega^*(\alpha^\downarrow, \beta^\uparrow)$. In
particular, $\Omega^*(\alpha^\downarrow, \beta^\uparrow)$ is
monotonically increasing in $G$. Investigating $G(y)$, it turns
out that $G(y)$ is in turn increasing in $y$. Since $y$ is bounded
from above by $y=0$,  the maximal value for $G(y)$ is provided by
$G(y=0)=1+\frac{1}{4}\sqrt{2}\ln{(3-2\sqrt{2})}$. Next, we note
that $\Omega^*(\alpha^\downarrow, \beta^\uparrow)$ is an
increasing function of $\sqrt{J}$. With the constraint that
$\alpha^\uparrow_{\textnormal{\scriptsize eff}}=\frac{1}{2}$, which is necessary for
$x_{\frac{1}{2}}=0$,  the largest $\sqrt{J}$ arises for
$\alpha^\downarrow=0$, i.e. $\sqrt{J}=1$. Combining both results,
the maximal value for the critical rates
$\Omega^*(\alpha^\downarrow, \beta^\uparrow)$ occurs for
$\alpha^\downarrow=0$ and $y=0$. Both conditions together yield
 \begin{equation}
\Omega_C=1+\frac{1}{4}\sqrt{2}\ln{(3-2\sqrt{2})} \,.
\end{equation}
Finally, we note that $\alpha^\downarrow=0$ and $y=0$ implies
$\beta^\uparrow=0$, such that $\Omega_C$ arises if
$\alpha^\downarrow=0$ and $\beta^\uparrow=0$.

\section*{References}
 
\bibliographystyle{unsrt}


\begin{thebibliography}{<11>}
\expandafter\ifx\csname natexlab\endcsname\relax\def\natexlab#1{#1}\fi
\expandafter\ifx\csname bibnamefont\endcsname\relax
  \def\bibnamefont#1{#1}\fi
\expandafter\ifx\csname bibfnamefont\endcsname\relax
  \def\bibfnamefont#1{#1}\fi
\expandafter\ifx\csname citenamefont\endcsname\relax
  \def\citenamefont#1{#1}\fi
\expandafter\ifx\csname url\endcsname\relax
  \def\url#1{\texttt{#1}}\fi
\expandafter\ifx\csname urlprefix\endcsname\relax\def\urlprefix{URL }\fi
\providecommand{\bibinfo}[2]{#2}
\providecommand{\eprint}[2][]{\url{#2}}

\bibitem{barabasi}
\bibinfo{author}{\bibfnamefont{A.}~\bibnamefont{Barabasi}} \bibnamefont{and}
  \bibinfo{author}{\bibfnamefont{H.}~\bibnamefont{Stanley}},
  \emph{\bibinfo{title}{Fractal Concepts in Surface Growth}}
  (\bibinfo{publisher}{Cambridge University Press, Cambridge, England},
  \bibinfo{year}{1995}).

\bibitem{deutscher}
\bibinfo{editor}{\bibfnamefont{G.}~\bibnamefont{Deutscher}},
  \bibinfo{editor}{\bibfnamefont{R.}~\bibnamefont{Zallan}}, \bibnamefont{and}
  \bibinfo{editor}{\bibfnamefont{J.}~\bibnamefont{Adler}}, eds.,
  \emph{\bibinfo{title}{Percolation Structures and Processes}},
  vol.~\bibinfo{volume}{5} of \emph{\bibinfo{series}{Annals of the Israel
  Physical Society}} (\bibinfo{publisher}{Adam Hilger, Bristol},
  \bibinfo{year}{1983}).

\bibitem{droz-1989-39}
\bibinfo{author}{\bibfnamefont{M.}~\bibnamefont{Droz}},
  \bibinfo{author}{\bibfnamefont{Z.}~\bibnamefont{R\'acz}}, \bibnamefont{and}
  \bibinfo{author}{\bibfnamefont{J.}~\bibnamefont{Schmidt}},
  \bibinfo{journal}{Phys. Rev. A} \textbf{\bibinfo{volume}{39}},
  \bibinfo{pages}{2141} (\bibinfo{year}{1989}).

\bibitem{mattis-1998-70}
\bibinfo{author}{\bibfnamefont{D.~C.} \bibnamefont{Mattis}} \bibnamefont{and}
  \bibinfo{author}{\bibfnamefont{M.~L.} \bibnamefont{Glasser}},
  \bibinfo{journal}{Rev. Mod. Phys.} \textbf{\bibinfo{volume}{70}},
  \bibinfo{pages}{979} (\bibinfo{year}{1998}).

\bibitem{SchmittmannZia}
\bibinfo{author}{\bibfnamefont{B.}~\bibnamefont{Schmittmann}} \bibnamefont{and}
  \bibinfo{author}{\bibfnamefont{R.}~\bibnamefont{Zia}}, in
  \emph{\bibinfo{booktitle}{Phase Transitions and Critical Phenomena}}, edited
  by \bibinfo{editor}{\bibfnamefont{C.}~\bibnamefont{Domb}} \bibnamefont{and}
  \bibinfo{editor}{\bibfnamefont{J.}~\bibnamefont{Lebowitz}}
  (\bibinfo{publisher}{Academic Press, London}, \bibinfo{year}{1995}),
  vol.~\bibinfo{volume}{17}.

\bibitem{macdonald-1968-6}
\bibinfo{author}{\bibfnamefont{C.}~\bibnamefont{MacDonald}},
  \bibinfo{author}{\bibfnamefont{J.}~\bibnamefont{Gibbs}}, \bibnamefont{and}
  \bibinfo{author}{\bibfnamefont{A.}~\bibnamefont{Pipkin}},
  \bibinfo{journal}{Biopolymers} \textbf{\bibinfo{volume}{6}},
  \bibinfo{pages}{1} (\bibinfo{year}{1968}).

\bibitem{Howard}
\bibinfo{author}{\bibfnamefont{J.}~\bibnamefont{Howard}},
  \emph{\bibinfo{title}{Mechanics of Motor Proteins and the Cytoskeleton}}
  (\bibinfo{publisher}{Sinauer Press, Sunderland, Massachusetts},
  \bibinfo{year}{2001}).

\bibitem{hirokawa-1998-279}
\bibinfo{author}{\bibfnamefont{N.}~\bibnamefont{Hirokawa}},
  \bibinfo{journal}{Science} \textbf{\bibinfo{volume}{279}},
  \bibinfo{pages}{519} (\bibinfo{year}{1998}).

\bibitem{helbing-2001-73}
\bibinfo{author}{\bibfnamefont{D.}~\bibnamefont{Helbing}},
  \bibinfo{journal}{Rev. Mod. Phys.} \textbf{\bibinfo{volume}{73}},
  \bibinfo{pages}{1067} (\bibinfo{year}{2001}).

\bibitem{chowdhury-2000-329}
\bibinfo{author}{\bibfnamefont{D.}~\bibnamefont{Chowdhury}},
  \bibinfo{author}{\bibfnamefont{L.}~\bibnamefont{Santen}}, \bibnamefont{and}
  \bibinfo{author}{\bibfnamefont{A.}~\bibnamefont{Schadschneider}},
  \bibinfo{journal}{Physics Reports} \textbf{\bibinfo{volume}{329}},
  \bibinfo{pages}{199} (\bibinfo{year}{2000}).

\bibitem{Derrida}
\bibinfo{author}{\bibfnamefont{B.}~\bibnamefont{Derrida}} \bibnamefont{and}
  \bibinfo{author}{\bibfnamefont{M.}~\bibnamefont{Evans}}, in
  \emph{\bibinfo{booktitle}{Nonequilibrium Statistical Mechanics in One
  Dimension}}, edited by
  \bibinfo{editor}{\bibfnamefont{V.}~\bibnamefont{Privman}}
  (\bibinfo{publisher}{Cambridge University Press, Cambridge},
  \bibinfo{year}{1997}), pp. \bibinfo{pages}{277--304}.

\bibitem{Mukamel}
\bibinfo{author}{\bibfnamefont{D.}~\bibnamefont{Mukamel}}, in
  \emph{\bibinfo{booktitle}{Soft and Fragile Matter}}, edited by
  \bibinfo{editor}{\bibfnamefont{M.}~\bibnamefont{Cates}} \bibnamefont{and}
  \bibinfo{editor}{\bibfnamefont{M.}~\bibnamefont{Evans}}
  (\bibinfo{publisher}{Institute of Physics Publishing, Bristol},
  \bibinfo{year}{2000}), pp. \bibinfo{pages}{237--258}.

\bibitem{Schutz}
\bibinfo{author}{\bibfnamefont{G.}~\bibnamefont{Sch\"utz}}, in
  \emph{\bibinfo{booktitle}{Phase Transitions and Critical Phenomena}}, edited
  by \bibinfo{editor}{\bibfnamefont{C.}~\bibnamefont{Domb}} \bibnamefont{and}
  \bibinfo{editor}{\bibfnamefont{J.}~\bibnamefont{Lebowitz}}
  (\bibinfo{publisher}{Academic Press, San Diego}, \bibinfo{year}{2001}),
  vol.~\bibinfo{volume}{19}, pp. \bibinfo{pages}{3--251}.

\bibitem{krug-1991-76}
\bibinfo{author}{\bibfnamefont{J.}~\bibnamefont{Krug}}, \bibinfo{journal}{Phys.
  Rev. Lett.} \textbf{\bibinfo{volume}{67}}, \bibinfo{pages}{1882}
  (\bibinfo{year}{1991}).

\bibitem{odor-2004-76}
\bibinfo{author}{\bibfnamefont{G.}~\bibnamefont{Odor}}, \bibinfo{journal}{Rev.
  Mod. Phys.} \textbf{\bibinfo{volume}{76}}, \bibinfo{pages}{663}
  (\bibinfo{year}{2004}).

\bibitem{tauber-1998-80}
\bibinfo{author}{\bibfnamefont{U.~C.} \bibnamefont{T\"auber}},
  \bibinfo{author}{\bibfnamefont{M.~J.} \bibnamefont{Howard}},
  \bibnamefont{and}
  \bibinfo{author}{\bibfnamefont{H.}~\bibnamefont{Hinrichsen}},
  \bibinfo{journal}{Phys. Rev. Lett.} \textbf{\bibinfo{volume}{80}},
  \bibinfo{pages}{2165} (\bibinfo{year}{1998}).

\bibitem{noh-2005-94}
\bibinfo{author}{\bibfnamefont{J.~D.} \bibnamefont{Noh}} \bibnamefont{and}
  \bibinfo{author}{\bibfnamefont{H.}~\bibnamefont{Park}},
  \bibinfo{journal}{Phys. Rev. Lett.} \textbf{\bibinfo{volume}{94}},
  \bibinfo{pages}{145702} (\bibinfo{year}{2005}).

\bibitem{dagotto-1995-271}
\bibinfo{author}{\bibfnamefont{E.}~\bibnamefont{Dagotto}} \bibnamefont{and}
  \bibinfo{author}{\bibfnamefont{T.~M.} \bibnamefont{Rice}},
  \bibinfo{journal}{Science} \textbf{\bibinfo{volume}{271}},
  \bibinfo{pages}{618} (\bibinfo{year}{1996}).

\bibitem{reichenbach-2006-97}
\bibinfo{author}{\bibfnamefont{T.}~\bibnamefont{Reichenbach}},
  \bibinfo{author}{\bibfnamefont{T.}~\bibnamefont{Franosch}}, \bibnamefont{and}
  \bibinfo{author}{\bibfnamefont{E.}~\bibnamefont{Frey}},
  \bibinfo{journal}{Phys. Rev. Lett.} \textbf{\bibinfo{volume}{97}},
  \bibinfo{pages}{050603} (\bibinfo{year}{2006}).

\bibitem{zutic-2004-76}
\bibinfo{author}{\bibfnamefont{I.}~\bibnamefont{Zutic}},
  \bibinfo{author}{\bibfnamefont{J.}~\bibnamefont{Fabian}}, \bibnamefont{and}
  \bibinfo{author}{\bibfnamefont{S.~D.} \bibnamefont{Sarma}},
  \bibinfo{journal}{Rev. Mod. Phys.} \textbf{\bibinfo{volume}{76}},
  \bibinfo{pages}{323} (\bibinfo{year}{2004}).

\bibitem{hahn-1998-73}
\bibinfo{author}{\bibfnamefont{C.~K.} \bibnamefont{Hahn}},
  \bibinfo{author}{\bibfnamefont{Y.~J.} \bibnamefont{Park}},
  \bibinfo{author}{\bibfnamefont{E.~K.} \bibnamefont{Kim}}, \bibnamefont{and}
  \bibinfo{author}{\bibfnamefont{S.}~\bibnamefont{Min}},
  \bibinfo{journal}{Appl. Phys. Lett.} \textbf{\bibinfo{volume}{73}},
  \bibinfo{pages}{2479} (\bibinfo{year}{1998}).

\bibitem{Hinsch}
\bibinfo{author}{\bibfnamefont{H.}~\bibnamefont{Hinsch}},
  \bibinfo{author}{\bibfnamefont{R.}~\bibnamefont{Kouyos}}, \bibnamefont{and}
  \bibinfo{author}{\bibfnamefont{E.}~\bibnamefont{Frey}}, in
  \emph{\bibinfo{booktitle}{Traffic and Granular Flow '05}}, edited by
  \bibinfo{editor}{\bibfnamefont{A.}~\bibnamefont{Schadschneider}},
  \bibinfo{editor}{\bibfnamefont{T.}~\bibnamefont{P\"oschel}},
  \bibinfo{editor}{\bibfnamefont{R.}~\bibnamefont{K\"uhne}},
  \bibinfo{editor}{\bibfnamefont{M.}~\bibnamefont{Schreckenberg}},
  \bibnamefont{and} \bibinfo{editor}{\bibfnamefont{D.~E.} \bibnamefont{Wolf}}
  (\bibinfo{publisher}{Springer}, \bibinfo{year}{2006}).

\bibitem{popkov-2001-64}
\bibinfo{author}{\bibfnamefont{V.}~\bibnamefont{Popkov}} \bibnamefont{and}
  \bibinfo{author}{\bibfnamefont{I.}~\bibnamefont{Peschel}},
  \bibinfo{journal}{Phys. Rev. E} \textbf{\bibinfo{volume}{64}},
  \bibinfo{pages}{026126} (\bibinfo{year}{2001}).

\bibitem{popkov-2003-112}
\bibinfo{author}{\bibfnamefont{V.}~\bibnamefont{Popkov}} \bibnamefont{and}
  \bibinfo{author}{\bibfnamefont{G.~M.} \bibnamefont{Sch\"utz}},
  \bibinfo{journal}{J. Stat. Phys} \textbf{\bibinfo{volume}{112}},
  \bibinfo{pages}{523} (\bibinfo{year}{2003}).

\bibitem{popkov-2004-37}
\bibinfo{author}{\bibfnamefont{V.}~\bibnamefont{Popkov}}, \bibinfo{journal}{J.
  Phys. A: Math. Gen.} \textbf{\bibinfo{volume}{37}}, \bibinfo{pages}{1545}
  (\bibinfo{year}{2004}).

\bibitem{pronina-2004-37}
\bibinfo{author}{\bibfnamefont{E.}~\bibnamefont{Pronina}} \bibnamefont{and}
  \bibinfo{author}{\bibfnamefont{A.~B.} \bibnamefont{Kolomeisky}},
  \bibinfo{journal}{J. Phys. A: Math. Gen.} \textbf{\bibinfo{volume}{37}},
  \bibinfo{pages}{9907} (\bibinfo{year}{2004}).

\bibitem{mitsudo-2005-38}
\bibinfo{author}{\bibfnamefont{T.}~\bibnamefont{Mitsudo}} \bibnamefont{and}
  \bibinfo{author}{\bibfnamefont{H.}~\bibnamefont{Hayakawa}},
  \bibinfo{journal}{J. Phys. A: Math. Gen.} \textbf{\bibinfo{volume}{38}},
  \bibinfo{pages}{3087} (\bibinfo{year}{2005}).

\bibitem{pronina-2006-372}
\bibinfo{author}{\bibfnamefont{E.}~\bibnamefont{Pronina}} \bibnamefont{and}
  \bibinfo{author}{\bibfnamefont{A.~B.} \bibnamefont{Kolomeisky}},
  \bibinfo{journal}{Physica A} \textbf{\bibinfo{volume}{372}},
  \bibinfo{pages}{12} (\bibinfo{year}{2006}).

\bibitem{reichenbach-deloc}
\bibinfo{author}{\bibfnamefont{T.}~\bibnamefont{Reichenbach}},
  \bibinfo{author}{\bibfnamefont{T.}~\bibnamefont{Franosch}}, \bibnamefont{and}
  \bibinfo{author}{\bibfnamefont{E.}~\bibnamefont{Frey}}, \bibinfo{note}{in
  preparation}.

\bibitem{shaw-2003-68}
\bibinfo{author}{\bibfnamefont{L.~B.} \bibnamefont{Shaw}},
  \bibinfo{author}{\bibfnamefont{R.~K.~P.} \bibnamefont{Zia}},
  \bibnamefont{and} \bibinfo{author}{\bibfnamefont{K.~H.} \bibnamefont{Lee}},
  \bibinfo{journal}{Phys. Rev. E} \textbf{\bibinfo{volume}{68}},
  \bibinfo{pages}{021910} (\bibinfo{year}{2003}).

\bibitem{pierobon-2006-74}
\bibinfo{author}{\bibfnamefont{P.}~\bibnamefont{Pierobon}},
  \bibinfo{author}{\bibfnamefont{T.}~\bibnamefont{Franosch}}, \bibnamefont{and}
  \bibinfo{author}{\bibfnamefont{E.}~\bibnamefont{Frey}},
  \bibinfo{journal}{Phys. Rev. E} \textbf{\bibinfo{volume}{74}},
  \bibinfo{pages}{031920} (\bibinfo{year}{2006}).

\bibitem{derrida-1998-301}
\bibinfo{author}{\bibfnamefont{B.}~\bibnamefont{Derrida}}, 
  \bibinfo{journal}{Phys. Rep.} \textbf{\bibinfo{volume}{301}},
  \bibinfo{pages}{65} (\bibinfo{year}{1998}).
  
  \bibitem{klumpp-2003-113}
\bibinfo{author}{\bibfnamefont{S.}~\bibnamefont{Klumpp}} \bibnamefont{and}
  \bibinfo{author}{\bibfnamefont{R.}~\bibnamefont{Lipowsky}},
  \bibinfo{journal}{J. Stat. Phys.} \textbf{\bibinfo{volume}{113}},
  \bibinfo{pages}{233} (\bibinfo{year}{2003}).
  
  \bibitem{parmeggiani-2003-90}
\bibinfo{author}{\bibfnamefont{A.}~\bibnamefont{Parmeggiani}},
  \bibinfo{author}{\bibfnamefont{T.}~\bibnamefont{Franosch}}, \bibnamefont{and}
  \bibinfo{author}{\bibfnamefont{E.}~\bibnamefont{Frey}},
  \bibinfo{journal}{Phys. Rev. Lett.} \textbf{\bibinfo{volume}{90}},
  \bibinfo{pages}{086601} (\bibinfo{year}{2003}).

\bibitem{parmeggiani-2004-70}
\bibinfo{author}{\bibfnamefont{A.}~\bibnamefont{Parmeggiani}},
  \bibinfo{author}{\bibfnamefont{T.}~\bibnamefont{Franosch}}, \bibnamefont{and}
  \bibinfo{author}{\bibfnamefont{E.}~\bibnamefont{Frey}},
  \bibinfo{journal}{Phys. Rev. E} \textbf{\bibinfo{volume}{70}},
  \bibinfo{pages}{046101} (\bibinfo{year}{2004}).

\end{thebibliography}

\end{document}